\documentclass[ reprint, amsmath, amssymb, aps, showkeys]{revtex4-1}
\pdfoutput=1

\usepackage{graphicx}
\usepackage{dcolumn}
\usepackage{bm}
\usepackage{caption}
\usepackage{subcaption}
\usepackage{color,soul}
\usepackage{hyperref}
\usepackage{textcomp}
\usepackage[section]{placeins}
\usepackage{soul}
\usepackage[usenames, dvipsnames]{xcolor}

\newcommand{\Nblocks} {N_b}
\newcommand{\Mmm} {M_{\mu\mu}}

\newcommand{\delphes} {{\sc{Delphes}}}
\newcommand{\pythia} {{\sc{Pythia}}}
\newcommand{\numpy} {{\sc{Numpy}}}

\newcommand{\Ftotal} {F_{\mathrm{total}}}

\DeclareMathOperator{\sgn}{sgn}

\soulregister\cite7
\soulregister\ref7
\soulregister\pageref7

\begin{document}

    \title{Bayesian Block Histogramming for High Energy Physics}%

    \author{Brian Pollack, Saptaparna Bhattacharya, and Michael Schmitt}
    \affiliation{%
     Northwestern University
    }%

    \date{\today}

    \begin{abstract}
        The Bayesian Block algorithm, originally developed for applications in astronomy, can be
        used to improve the binning of histograms in high energy physics (HEP).  The visual improvement
        can be dramatic, as shown here with multiple examples.  More importantly, this algorithm
        produces histograms that accurately represent the underlying distribution while being robust to
        statistical fluctuations.  The algorithm is compared with other binning
        methods in a quantitative manner using distributions commonly found in HEP\@.
        These examples show the usefulness of the binning provided by the Bayesian Blocks algorithm
        both for presentation and modeling of data.
    \end{abstract}

    \keywords{High Energy Physics; Data Analysis; Bayesian Methods}

    \maketitle

\section{\label{sec:intro}Introduction}
\par
Histograms are ubiquitous in particle physics, yet histogram binning is usually
settled in an \emph{ad~hoc} manner.  Most of the time, a subjectively natural
range and bin width is chosen, motivated mainly by obtaining a nice-looking plot.
Objective methods have been proposed~\cite{Scott,Freedman,Gregory,Izenman,Knuth,Shimazaki} 
that determine binning according to some optimization procedure.  
For example, Scott's Rule~\cite{Scott} and the Freedman-Diaconis 
Rule~\cite{Freedman} determine the number of fixed-width bins by the number
of entries and a measure of the spread of the distribution (root-mean-square 
for Scott's Rule and interquartile range for Freedman-Diaconis Rule).  
Knuth's Rule~\cite{Knuth}
takes the structure of the distribution into account but uses bins of fixed width.  The method of
``equal population'' requires that each bin have similar numbers of entries, and thus the bin widths
may vary, but the location of the bin edges is chosen arbitrarily.
The Bayesian Block algorithm, in contrast, allows the bin widths to vary and determines the
bin edges based on the structure of the distribution.
\par
The Bayesian Blocks algorithm was developed in an astronomy context by 
Scargle~\cite{Scargle,Scargle98}.
His objective was to set bin edges, called ``change points'', at times when the light
flux from an astrophysical object suddenly changed.   The flux is represented by
the arrival times $t_n$ of photons in a telescope; given a set of event data $\{ t_n \}$ for 
$n = 1, \dots, N$, the algorithm uniquely determines the number and placement of 
the change points.   
\par
In our analysis, the change points play the role of histogram bin boundaries for a set of event data
$\{ x_n \}$. The resulting histogram is objective rather than subjective.  Ranges in $x$ in which
the data are sparse result in larger bins, and ranges in which the data are concentrated result in
smaller bins.  Furthermore, if the (empirical) probability density function (pdf) changes slowly,
then the bins are wide, and if it changes rapidly, the bins will be narrower.  The prospect of a
self-adjusting histogram is attractive especially in contexts in which the distributions falls over
orders of magnitude: typical histograms plotted on a semi-logarithmic scale either lose the
structure at the high values of the pdf or are plagued by statistical fluctuations in the tails.
Sometimes researchers employ unequal binning but the bins are still chosen in an arbitrary manner,
and the results are seldom completely satisfactory.
\par
We apply the Bayesian Blocks algorithm to histogramming in collider physics.  We provide
illustrations of how this algorithm produces clear and visually pleasing histograms with minimal
subjective input from the analyzer.  We do not consider external factors that motivate histogram bin
width (e.g.\ resolution of data, systematic errors, comparison with existing histograms).  In
principle, these factors could be incorporated as additional inputs or constraints to a binning
algorithm, but will not be explored in this analysis. Naturally, the Bayesian Block algorithm can be
applied to any scientific field in which histograms are employed.
\par
Beyond producing pleasing histograms, we have discovered that the binning provided by the Bayesian
Blocks algorithm accurately represents the underlying distribution of the given data set while
suppressing the effects of statistical fluctuations.  Since the binning is optimal such that each
bin is consistent with a flat distribution, a given binning is robust to change when compared with
statistically independent data sets produced from the same underlying pdf.  We give multiple
examples to illustrate these points, all of which are inspired from common scenarios in HEP (high
energy physics) and span large ranges of bin occupancy.
\par
The remainder of this paper is structured as follows.  A brief technical description of the Bayesian
Blocks algorithm is given in Section~\ref{sec:description} followed by illustrations from collider
physics in Section~\ref{sec:vis_examples}.  Section~\ref{sec:bin_comp} compares the Bayesian Block
algorithm with several  other binning heuristics using quantitative metrics.  We conclude in
Section~\ref{sec:conclusion}.
\section{\label{sec:description}The Bayesian Block Algorithm}
\par
We will briefly describe the Bayesian Block algorithm.  The complete detailed account of the
mathematical derivation and algorithm implementation can be found in Scargle~\cite{Scargle}.
The Bayesian Blocks algorithm is a nonparametric modeling technique for determining the optimal
segmentation of a given set of univariate random variables. Each block (or bin, in the context of
histograms) is consistent with a pdf with compact support; the entire dataset is represented by this 
collection of finite pdfs. For this analysis (and for most current implementations of Bayesian
Blocks), each pdf is a uniform distribution, which thereby defines the `Piecewise Constant Model'
as discussed in Ref.~\cite{Scargle}. The number of blocks and the edges of the blocks are determined
through optimization of a `fitness function', which is essentially a goodness-of-fit statistic
dependent only on the input data and a regularization parameter (discussed below).
\par
The set of blocks is gapless and non-overlapping, where the first block edge is defined by the first
data point, and the last block edge is defined by the last data point.  A block can contain between
1 and $N$ data points, where the sum of the contents of all the blocks must equal $N$. The algorithm
relies on the additivity of the fitness function, and thus the fitness of a given set of blocks is
equal to the sum of the fitnesses of the individual blocks. The total fitness, $\Ftotal$ for a given
dataset is:
\begin{equation} \label{eq:total_fitness_gen}
    \Ftotal = \sum_{i=1}^K f(B_i),
\end{equation}
where $f(B_i)$ is the fitness for an individual block, and $K$ is the total number of blocks.  The
additivity requirement of the total fitnesses allows the Bayesian Blocks algorithm to greatly
improve the execution time with respect to a brute-force method~\cite{Scargle}.
\par
Given an ordered set of $N$ data points, the algorithm determines the optimal set of $K+1$
change-points (and therefore $K$ blocks) by iterating through the data points, and caching the
current maximum fitness values and corresponding indices. For example, during iteration $n$ (where
data point $n$ is being evaluated), the potential total fitnesses are calculated as:
\begin{equation} \label{eq:temp_fitness}
    \Ftotal(n, m) = F_{m} + f(B_{m}^n), m=1,2,\ldots, n-1
\end{equation}
where $F_{m}$ is the optimal fitness as determined during iteration $m$, and $f(B_{m}^n)$ is
the fitness of the block bound between data points $n$ and $m$.  This potential total fitness is
calculated $n-1$ times at each iteration, and the maximum of those fitnesses along with the relevant
change-points are stored and used during the subsequent iterations. After the final iteration, $N$,
the change-points associated with the maximum total fitness are returned.  This method guarantees that the
global maximum fitness is obtained in $O(N^2)$, which is much more efficient than an
exhaustive search of all $2^N$ potential configurations.

For a series of discrete, independent events, the fitness function for an individual block, $f(B_i)$, can
be defined as an unbinned log-likelihood (the so-called Cash statistic~\cite{Cash}):
\begin{equation} \label{eq:cash_block}
    f(B_i) = \ln{L_{i}(\lambda)} = N_{i}\ln\lambda-\lambda h_{i}.
\end{equation}
This modified Cash statistic is derived from the Poisson likelihood of $N$ events sampled from a
model with amplitude $\lambda$ over a range $h$.  It follows that the total fitness $\Ftotal$ is:
\begin{equation} \label{eq:total_fitness_cash}
    \Ftotal = \sum_{i=1}^{K}\ln{L_{i}^{\max}(\lambda)}.
\end{equation}

The fitness described above must be modified by a penalty term for the number of blocks.  Without
explicitly adding this additional parameter, there is an implicit assumption of a uniform prior on
the number of blocks between 0 and $N$. This is unreasonable in most cases, as typically 
$\Nblocks \ll N$, where $\Nblocks$ is the number of blocks. 
In Ref.~\cite{Scargle} a geometric prior of the form was chosen:
\begin{equation} \label{eq:geo_prior}
    P(\Nblocks) = P_0 \, \gamma^{\Nblocks},
\end{equation}
where $\gamma$ is the single free parameter, and $P_0$ is a normalization constant.
This prior must be tuned in order to achieve a reasonable binning for a given dataset.  An
overly conservative value will suppress the detection of true change-points, while too liberal a
value will lead to spurious change points (eventually reaching the limit of $\Nblocks = N$). 
When $\gamma$ is less than one, it is the factor by  which $N$ blocks are favored over $N+1$
blocks. The prior can be interpreted as a control on the false-positive rate for detecting
change-points.  
\par
The prior can be determined empirically as a function of the false-positive rate through simple toy
studies, or through other means of optimization.  As will be explained in~\ref{sec:bin_comp},
the prior value for this analysis is determined by optimizing metrics used for quantitative
histogram comparison.  In general, the number of change-points is insensitive to a large range
of reasonable values for~$\lambda$.
\par
The code used to implement Bayesian Blocks and generate the histograms in this paper is located in
the Scikit-HEP python package~\cite{Scikit-HEP} and was written and maintained by one of the authors
[BP].
\section{\label{sec:vis_examples}Illustrations}
\par
\begin{figure*}[!htb]
    \centering
    \begin{subfigure}{\columnwidth}
        \centering
        \includegraphics[width=\columnwidth]{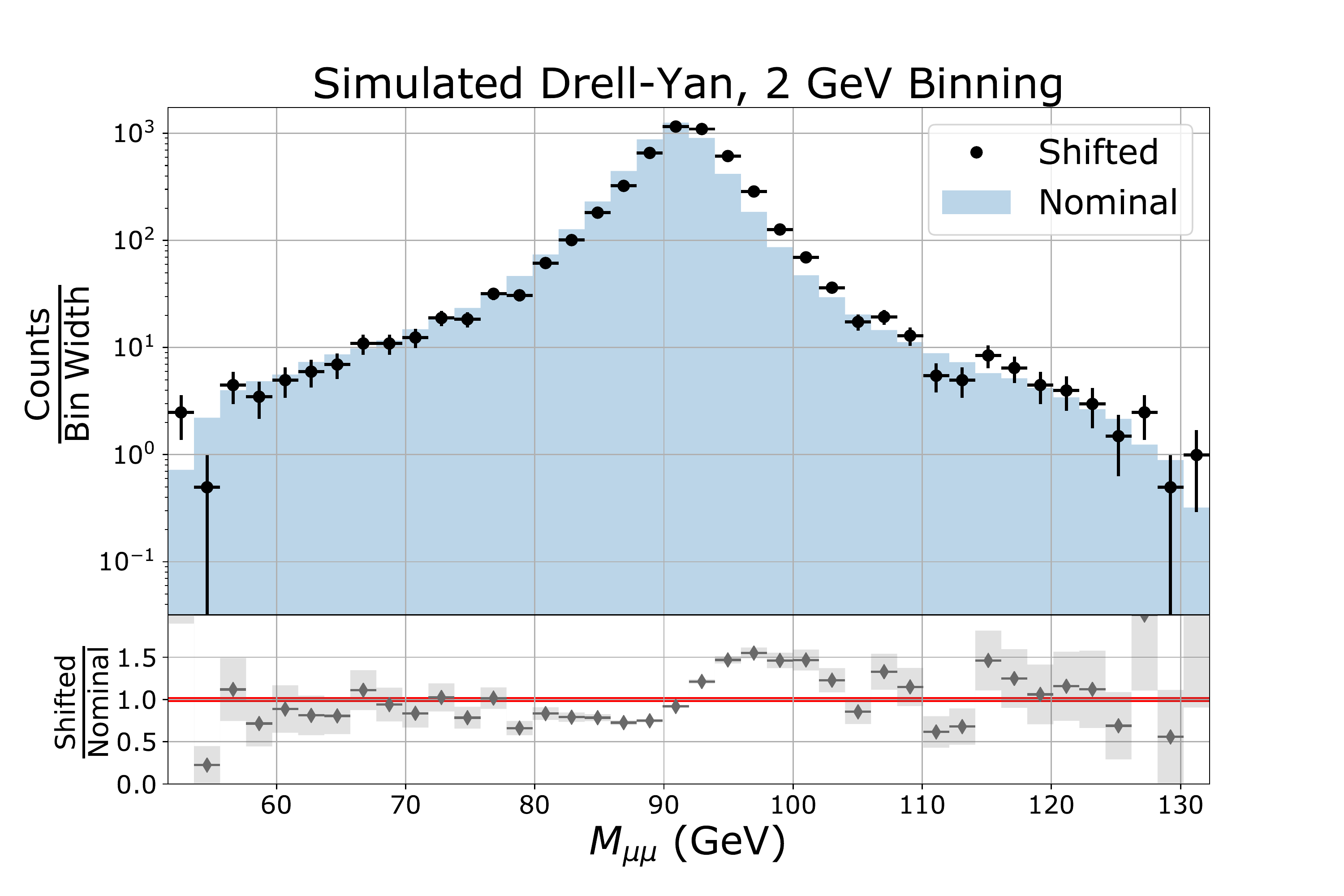}
        \caption{Fixed-width binning.\label{fig:dimuon-sub1}}
    \end{subfigure}
    \begin{subfigure}{\columnwidth}
        \centering
        \includegraphics[width=\columnwidth]{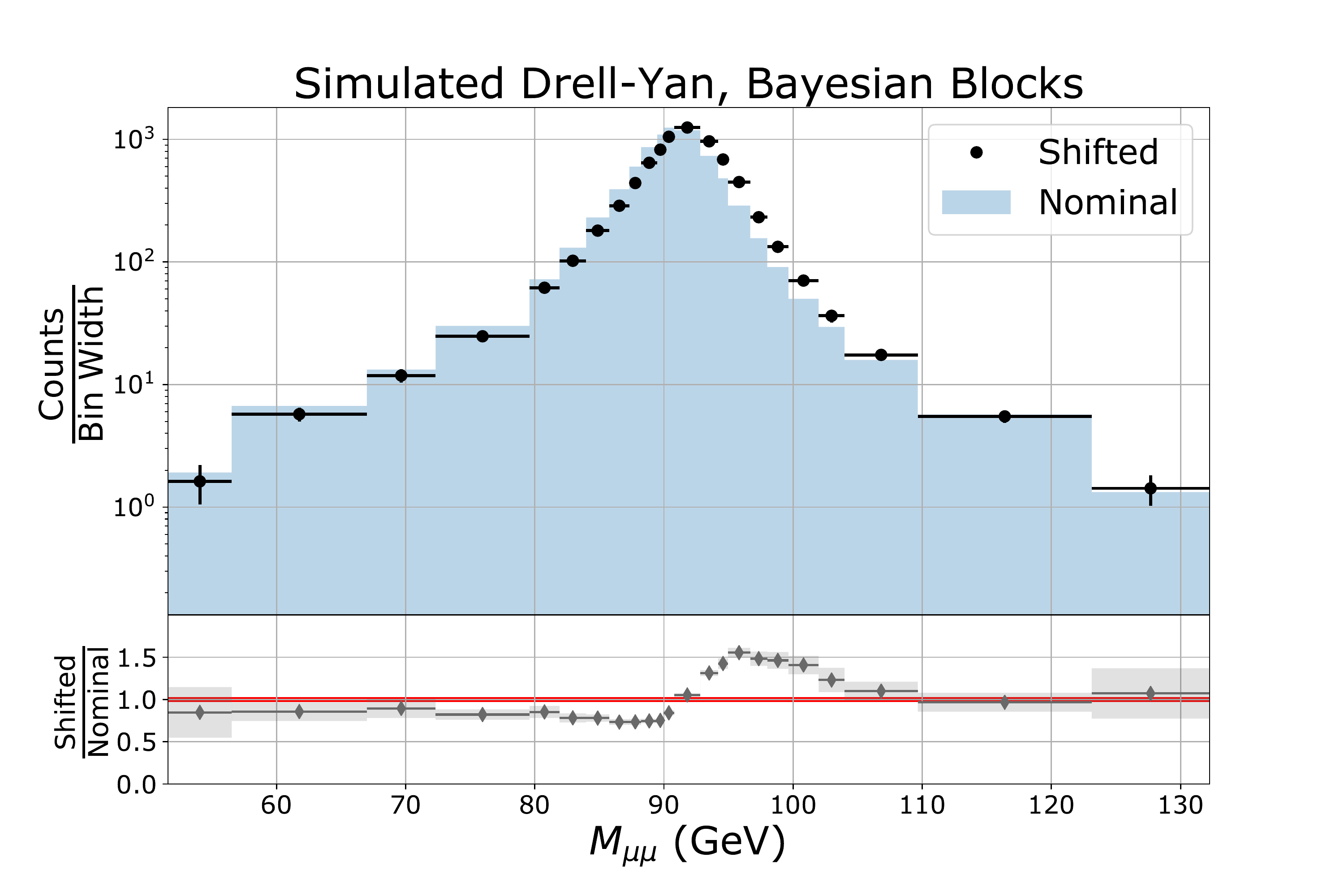}
        \caption{Bayesian Block binning.\label{fig:dimuon-sub2}}
    \end{subfigure}
    \caption{Comparison of simulated Drell-Yan distributions.\label{fig:dimuon}}
\end{figure*}

Our first illustration is the sharp peak in the distribution of the invariant
mass $\Mmm$ of muon pairs produced at a hadron collider; this resonance is the
Z~boson.   The example shown in Fig.~\ref{fig:dimuon} is produced using
simulated events, generated using the \pythia~\cite{pythia} and
\delphes~\cite{delphes} software packages.  In a typical application one
compares collider data (represented by black dots with error bars) to a
simulation (represented here by the light-blue histogram) expecting to see good
agreement.  If, for example, the momentum scale calibration for the data is not
quite correct, one will observe a shift in one distribution with respect to the
other.  Consequently, the ratio of the two histograms will display a
characteristic S-shape.   The clarity of the ratio of the two histograms is of
central importance in this type of diagnostic study.  For the sake of this
illustration, we have one set of simulated data with no modification, and a
second, independent sample in which the invariant mass values are shifted
by~1\%.  There are 10,000 events in the ``data'' sample, and $\approx 680,000$
events in the ``simulation'' sample.
\par
The left plot in Fig.~\ref{fig:dimuon} shows a typical choice of binning (2 GeV
width bins), and the right plot shows the binning obtained with the Bayesian
Block algorithm.  Each plot shows the distribution on a logarithmic scale and
the ratio of the two histograms on a linear scale.  The shaded regions show the
statistical uncertainties for the ratio.  The standard plot is unsatisfactory
because the statistical fluctuations below $\Mmm \approx 60$~GeV and above
$\Mmm \approx 125$~GeV are too large to allow any conclusions to be drawn about
the tails of these distributions.  Furthermore, the fairly sharp shape of the
peak near 90~GeV is not as clear and the ratio plot has a less pronounced
S-shaped curve.  The Bayesian Block plot, in contrast, shows a sharp peak and a
very clear S-shape, and the statistical fluctuations in the tails are greatly
reduced.  Since the widths of the Bayesian Blocks plot are not uniform, we
normalize each bin by its width.  We also normalize the standard histogram by
bin width.  The Bayesian Block algorithm produced 23 bins, whereas we used 40
bins for the standard histogram.  In this illustration, the Bayesian-Block
algorithm produces a superior visualization of the distribution and of the
differences between the two samples.  While this statement is qualitative, we
will show in the following section that the Bayesian Block algorithm is
quantitatively superior to other binning schemes in many cases.

Our second illustration is the distribution of the transverse momentum~$p_T$ of
a reconstructed jet produced in association with a vector boson; this
distribution is known to fall rapidly as $p_T$ increases and is characterized
by a long, sparsely-populated, high-energy tail.  High energy physicists will
look for new physics in tails like this.  Comparisons of data and Monte Carlo
simulations can be unsatisfactory when a uniform binning is employed.
\par
\begin{figure*}[!htb]
    \centering
    \begin{subfigure}{\columnwidth}
        \centering
        \includegraphics[width=\columnwidth]{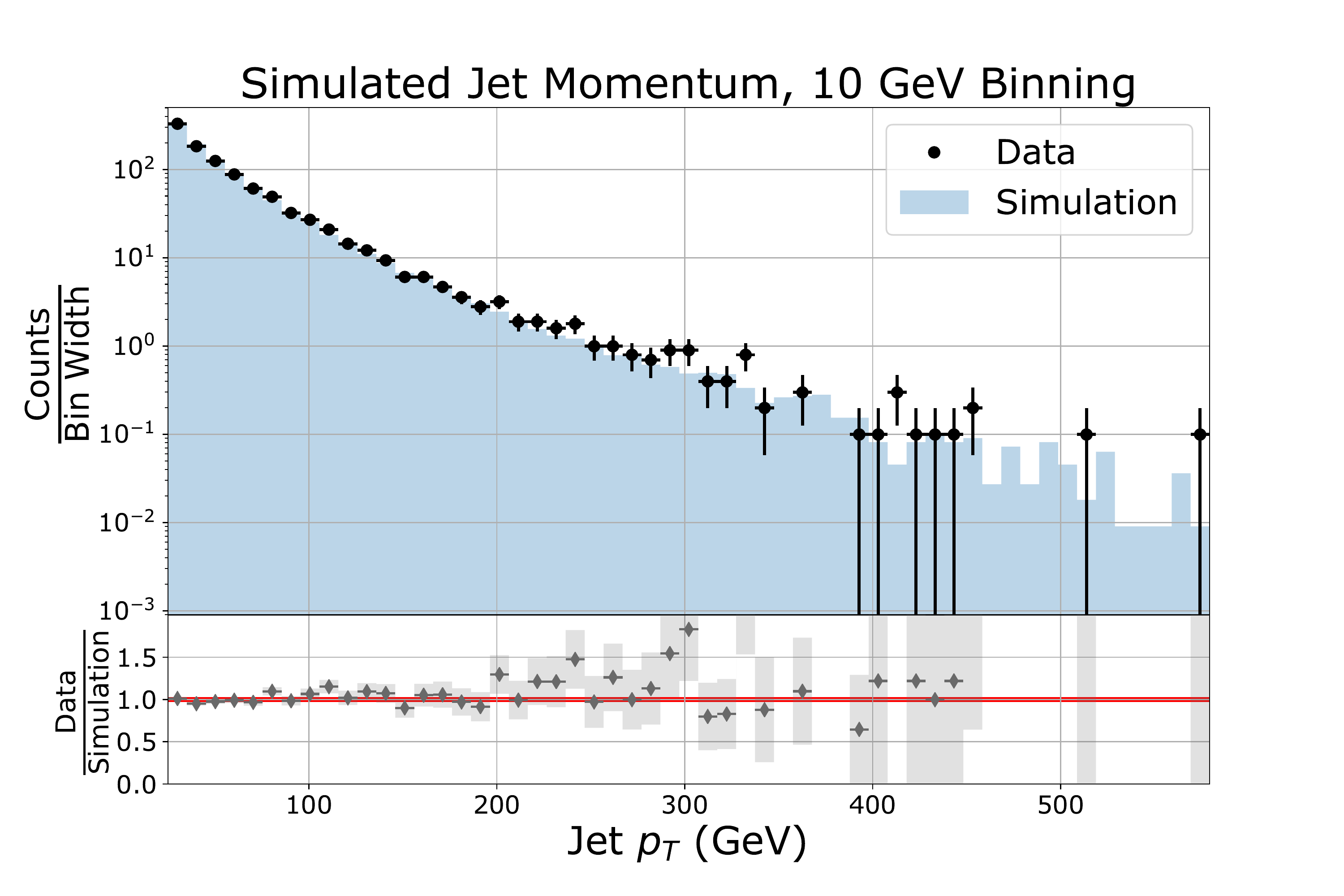}
        \caption{Fixed-width binning.\label{fig:jet-sub1}}
    \end{subfigure}
    \begin{subfigure}{\columnwidth}
        \centering
        \includegraphics[width=\columnwidth]{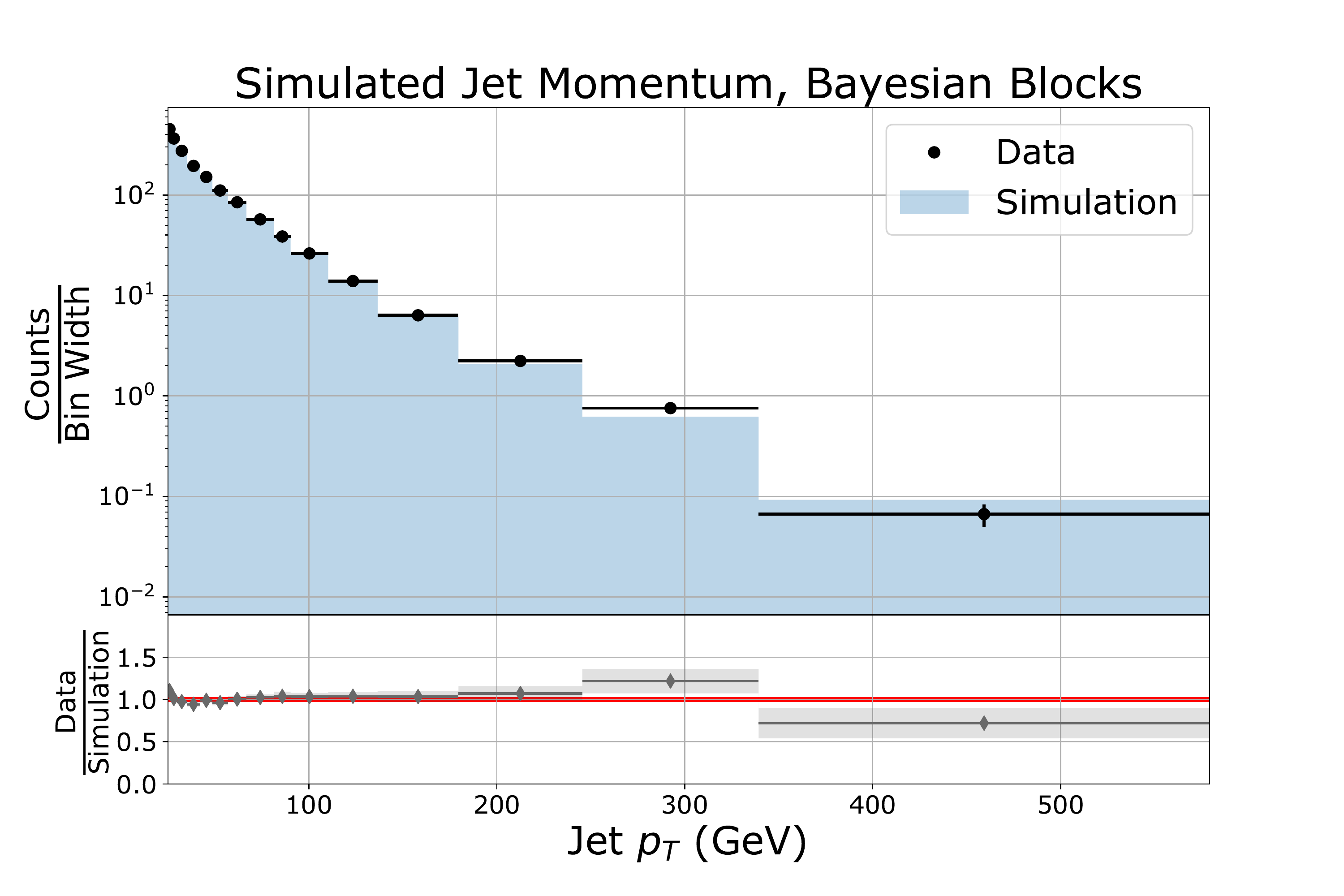}
        \caption{Bayesian Block binning.\label{fig:jet-sub2}}
    \end{subfigure}
    \caption{Comparison of simulated jet momentum distributions.\label{fig:jet}}
\end{figure*}

The two log plots in Fig.~\ref{fig:jet} show histograms produced with a typical
10 GeV binning and a binning determined by the Bayesian Blocks algorithm.
The uniformly binned histogram is reasonable in the low-momentum region.
However, it obscures an interesting minor disagreement between the data with the simulation that
occurs in the lowest momentum region.  Conversely, the high-momentum region is binned too finely,
and the ratio plot (lower panel) in the lower panel is difficult to interpret due to large
statistical uncertainties.  The Bayesian Block histogram suffers from none of these defects, again
producing a much more instructive and visually appealing plot.  This distribution will also be
examined quantitatively in the following section.
\par
These two examples serve to showcase the Bayesian Block algorithm in realistic HEP plotting
scenarios.  Our statements thus far are qualitative and subjective in nature, as is typically the
case when colloquially discussing the merits of data visualization methods.  However, the following
section will examine the Bayesian Block algorithm with a host of other common binning
schemes in order to objectively compare and evaluate performance.

\section{\label{sec:bin_comp}Comparison with Existing Methods}
\par
Many heuristics currently exist in order to generate an `optimal' histogram,
given an unbinned dataset.  However, the criteria for determining which binning
heuristic is optimal is not universally agreed upon.  Scott~\cite{Scott} and
others use the Integrated Mean Square Error (IMSE) to determine an optimal bin
width.  In practice, it is often difficult to obtain or approximate the
underlying pdf for a given distribution, which is needed to calculate the
IMSE\@.  Additionally, the minimization of the IMSE does not take into account
the visual appeal of the histogram to a user, which is an important, albeit
difficult to quantify, aspect of data visualization.
\par
In this section, we will compare multiple histogram binning methods with Bayesian Blocks, using
metrics inspired by the studies conducted by Lolla and Hoberock~\cite{lolla_hoberock}.  These
metrics will measure the visual appeal of a histogram and its ability to accurately reproduce the
underlying pdf, without overemphasising statistical fluctuations inherent in any given dataset.  The
motivations and definitions of these metrics will be covered in Section~\ref{sec:metrics}.
\par
We will evaluate the histograms using multiple different datasets with varying numbers of entries,
motivated by distributions commonly seen in HEP\@.  The two metric values will be displayed for all
histogram methods employed for each example case, along with the final combined rank for each
histogram method.  The following histogram methods will be examined, along with the Bayesian Blocks
algorithm:

\begin{enumerate}
    \item Sturges~\cite{Sturges}: $K=\lceil\ln(N)\rceil+1$.
    \item Doane~\cite{Doane}: Sturges rule, with a skewness correction.
    \item Scott~\cite{Scott}: $h=\frac{3.5\hat{\sigma}}{N^{1/2}}$, where $\hat{\sigma}$ is the standard deviation of the data.
    \item Freedman-Diaconis~\cite{Freedman}: $h=2\frac{\mathrm{IQR}i}{N^{1/3}}$, where IQR is the
        interquartile range of the data.
    \item Knuth~\cite{Knuth}: Fixed-width binning determined via likelihood minimization.
    \item Rice~\cite{Rice}: $K=\lceil2N^{1/3}\rceil$
    \item Square root: $K=\lceil\sqrt{N}\rceil$
    \item Equal population: Variable-width bins all contain equal number of entries.
\end{enumerate}

\subsection{\label{sec:metrics}Histogram Metrics}
\par
\begin{figure*}[!htb]
    \centering
    \includegraphics[width=\textwidth]{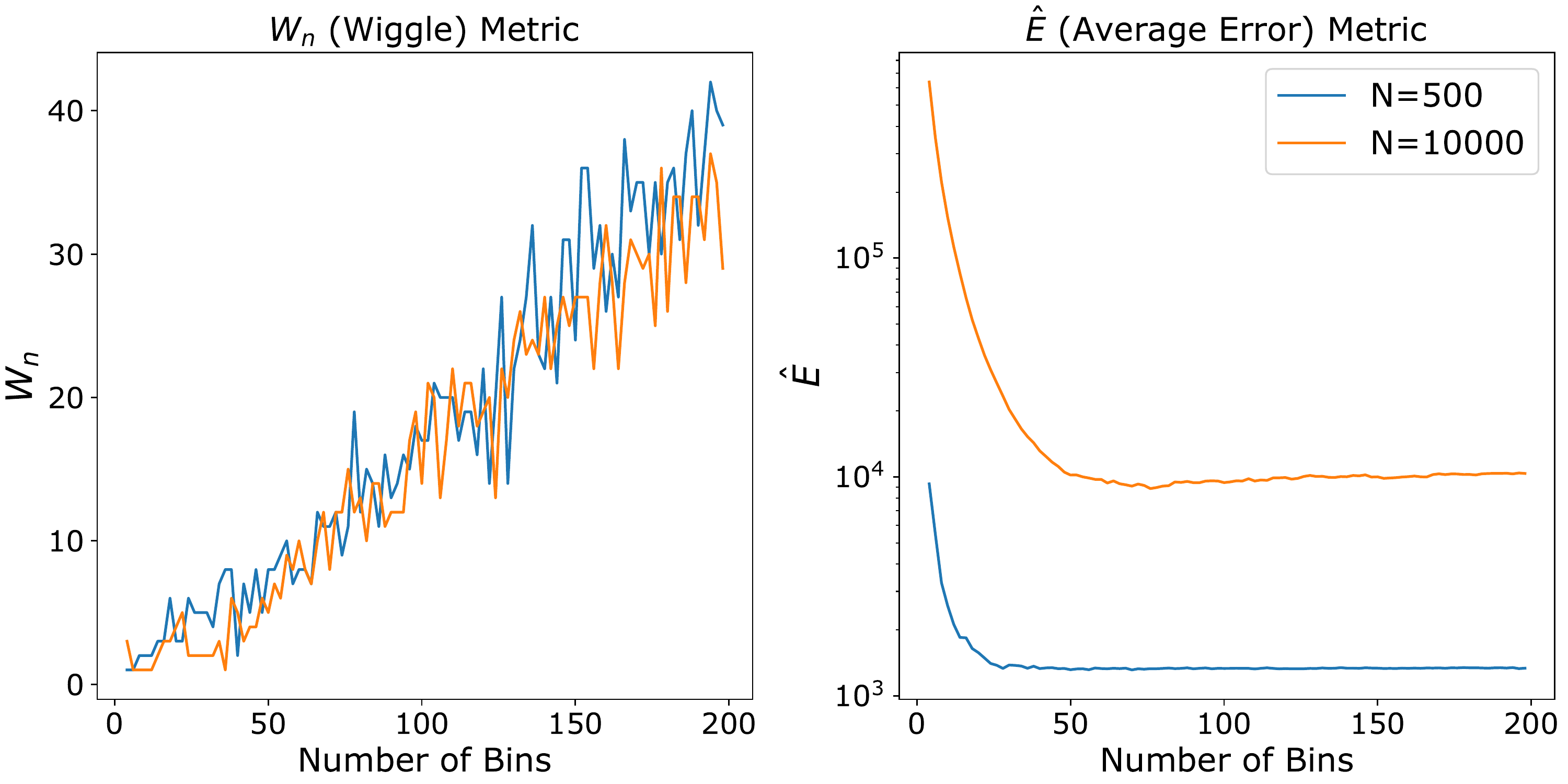}
    \caption{Metrics $W_n$ (left) and $\hat{E}$ (right) as a function of the number of (uniform
    width) bins. Distribution is jPT, for 500 and 10000 events.\label{fig:metric_example}}
\end{figure*}

Two metrics will be used to evaluate the performance of the
histogram binning methods.  The first metric is designed to capture the visual appeal of the
histogram by minimizing the number of bin-to-bin height fluctuations.  When dealing with a relatively
smooth underlying distribution, these features, or ``wiggles'', indicate that a histogram is picking
up the unwanted statistical fluctuations inherent in a finite dataset.  The number of wiggles in a
histogram is defined as: 
\begin{equation} \label{eq:wiggles}
    W_n = \sum[\sgn\left(f'(B_i)\right)\sgn\left(f'(B_{i+1})\right)=-1]
\end{equation}
where $f'(B_i)$ is the finite first derivative of the function describing the height of block (or
bin) $i$.  This metric simply counts the number of adjacent opposite-sign first derivatives, and
increases when there are many fluctuations in height from one bin to another.
A plot of $W_n$ as a function of the number of bins is shown in
Fig.~\ref{fig:metric_example}.
\par
The second metric measures the accuracy of a given histogram in reconstructing the underlying pdf,
while minimizing the impact of statistical fluctuations due to the initial data used to generate the
histogram.  Consider a dataset that consists of $N$ independent events, such that $D = \{d_1, d_2,
d_3,\ldots,d_N\}$.  From dataset $D$, construct a histogram.  From that histogram, generate a new
set of data $\hat{D} = \{\hat{d}_1, \hat{d}_2, \hat{d}_3,\ldots,\hat{d}_N\}$, where each data point
is generated by a linear interpolation of each bin (e.g.\ for a bin ranging from 0 to 1 in x with
height 5, we generate 5 equally spaced datum in x with values from 0 to 1.)  This results in a set
of data points equal in size to the original set, but evenly distributed within each respective bin.
One can compare the interpolated dataset, $\hat{D}$ with $M$ different independent datasets, all
of which are derived from the original distribution and have size $N$.  We can use these datasets to
construct the average error metric, defined as:
\begin{equation} \label{eq:avg_err}
    \hat{E} = \frac{1}{M}\sum_{m=1}^M\left(\sum_{n=1}^N|d_{nm}-\hat{d}_n|\right)
\end{equation}
where $d_{nm}$ is the $n$th data point from the $m$th data set.  This metric typically decreases as
the size of the bins decrease, but in general does not approach 0 as the bins become infinitesimally
narrow.  The metric penalizes a histogram for modeling the statistical fluctuations of a given
distribution by comparing the interpolated data with statistically independent datasets, and not the
dataset used to generate the histogram.  For the analyses performed in this manuscript, $M=100$
unless otherwise noted. A plot of $\hat{E}$ as a function of the number of bins is shown in
figure~\ref{fig:metric_example}.
\par
There is no obvious way to combine metrics Eq.~\ref{eq:wiggles} and Eq.~\ref{eq:avg_err} to produce an
overall metric.  However, we will show the results of the metrics in a two dimensional plane,
and also display the combined ranks of each histogram method.  For nine different histogram methods,
the best combined rank would be 2 (lowest relative score for each metric), and the worst combined
rank would be 18 (highest relative score for each metric).

\subsection{\label{sec:comparison}Comparison Results}
\par
This section will show the outputs and metric results of histograms generated with the
aforementioned binning heuristics for different distributions, and varying numbers of entries.  The
distributions examined are as follows:
\begin{enumerate}
    \item A Drell-Yan invariant mass distribution (DY).
    \item The transverse momentum of the leading muon from a Drell-Yan process (MuPT).
    \item The transverse momentum of a jet produced in association with a vector boson (jPT).
    \item A Gaussian distribution (Gauss).
    \item A bimodal distribution formed from two Laplace distributions on a uniform background
        (2LP).
\end{enumerate}
\par
The DY, MuPT, and jPT distributions are all simulated using the \pythia\ and \delphes\ simulation
software packages.  The Gaussian and bimodal distributions are generated using the python
package \numpy~\cite{numpy}.  Histograms are produced for each distribution with datasets of size
$N$=500, 1000,
5000, and 10000. Both the Bayesian Block and equal population binning methods have a user-defined
adjustable parameter.  The Bayesian Blocks prior (Eq.~\ref{eq:geo_prior}) and the number of bins in
the equal population method are determined by minimizing the combined metric ranks.
\par
\begin{figure*}[!htb]
    \centering
    \begin{subfigure}{\columnwidth}
        \centering
        \includegraphics[width=0.9\columnwidth]{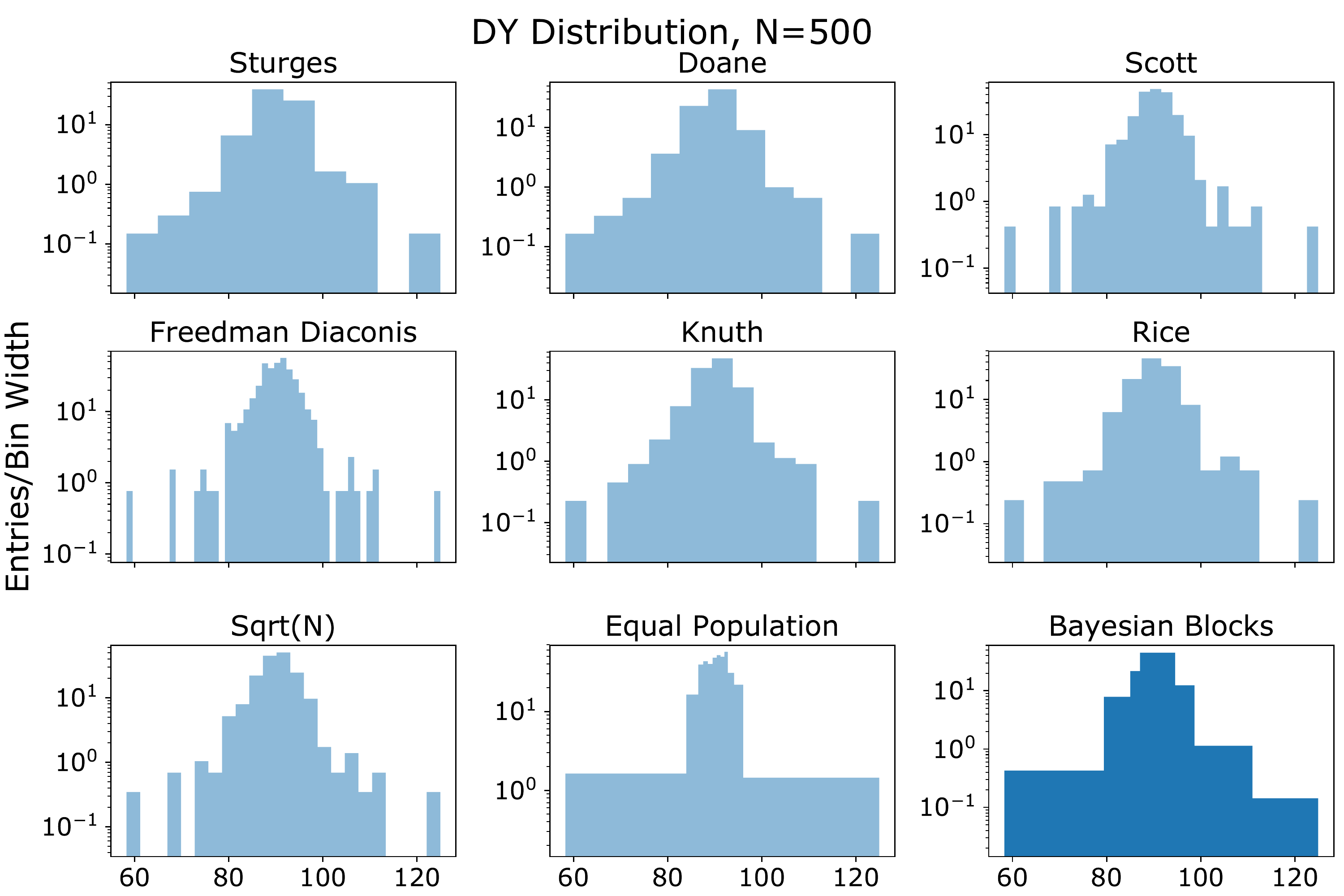}
        \vspace*{2mm}
    \end{subfigure}
    \begin{subfigure}{\columnwidth}
        \centering
        \includegraphics[width=0.9\columnwidth]{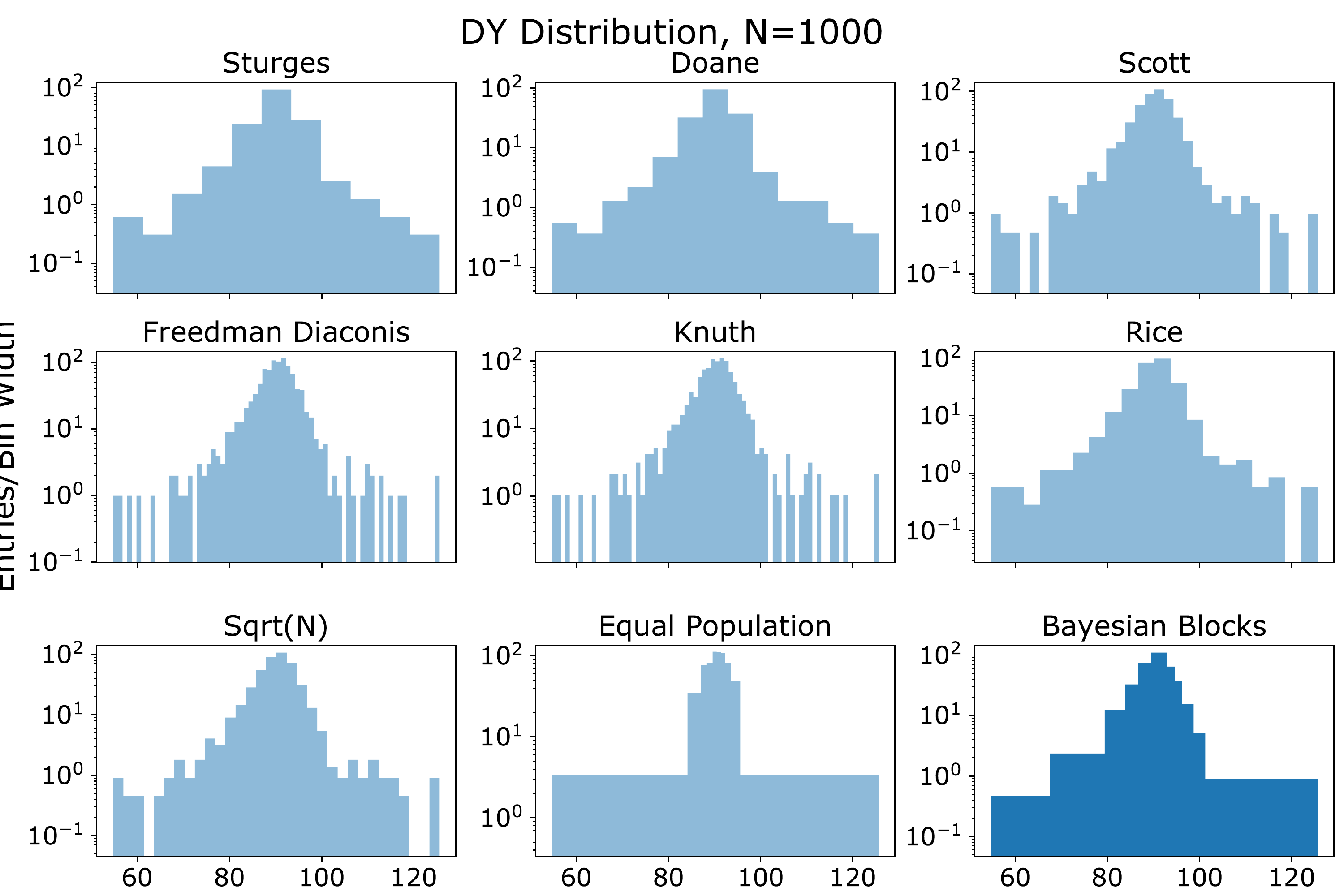}
        \vspace*{2mm}
    \end{subfigure}
    \begin{subfigure}{\columnwidth}
        \centering
        \includegraphics[width=0.9\columnwidth]{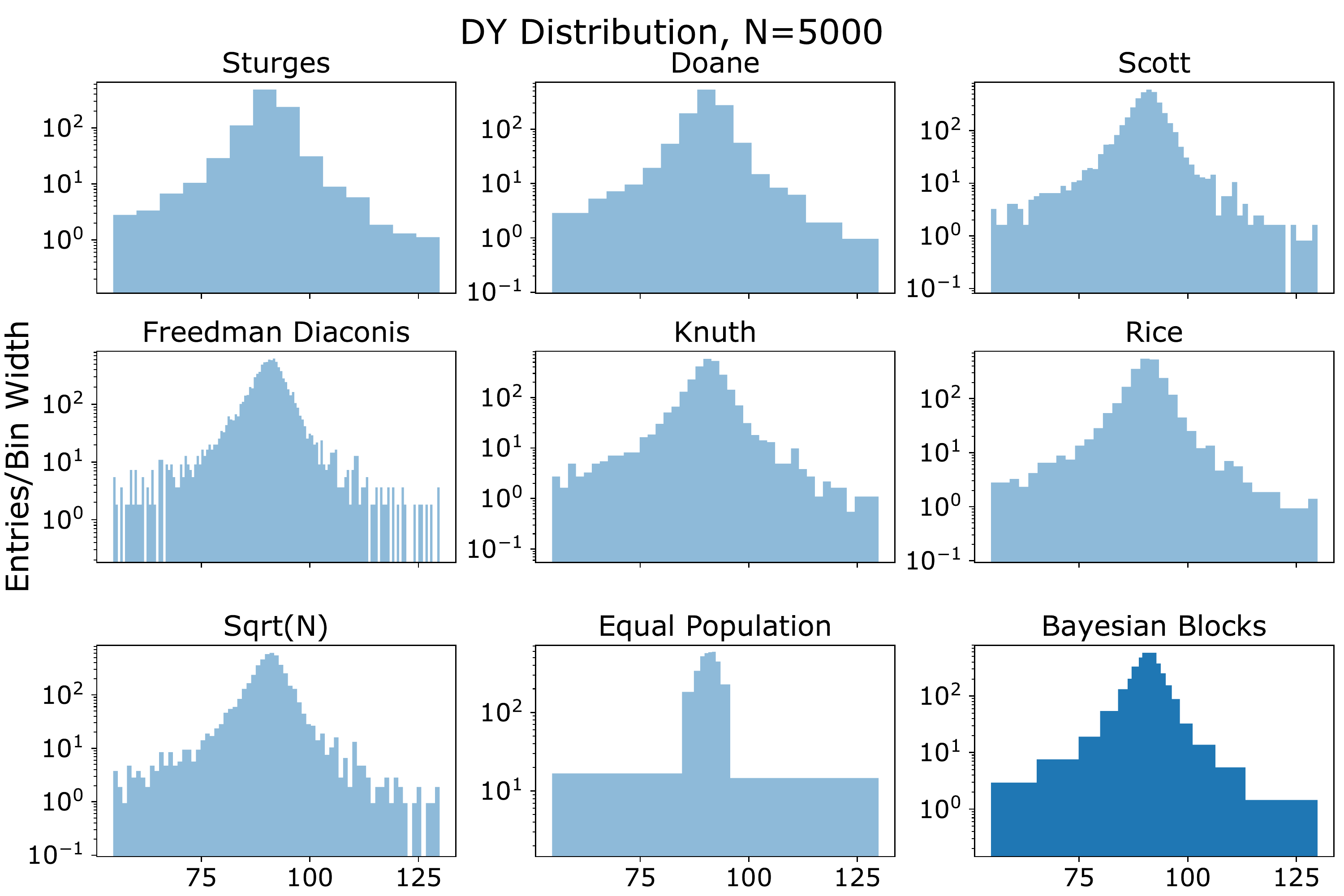}
        \vspace*{2mm}
    \end{subfigure}
    \begin{subfigure}{\columnwidth}
        \centering
        \includegraphics[width=0.9\columnwidth]{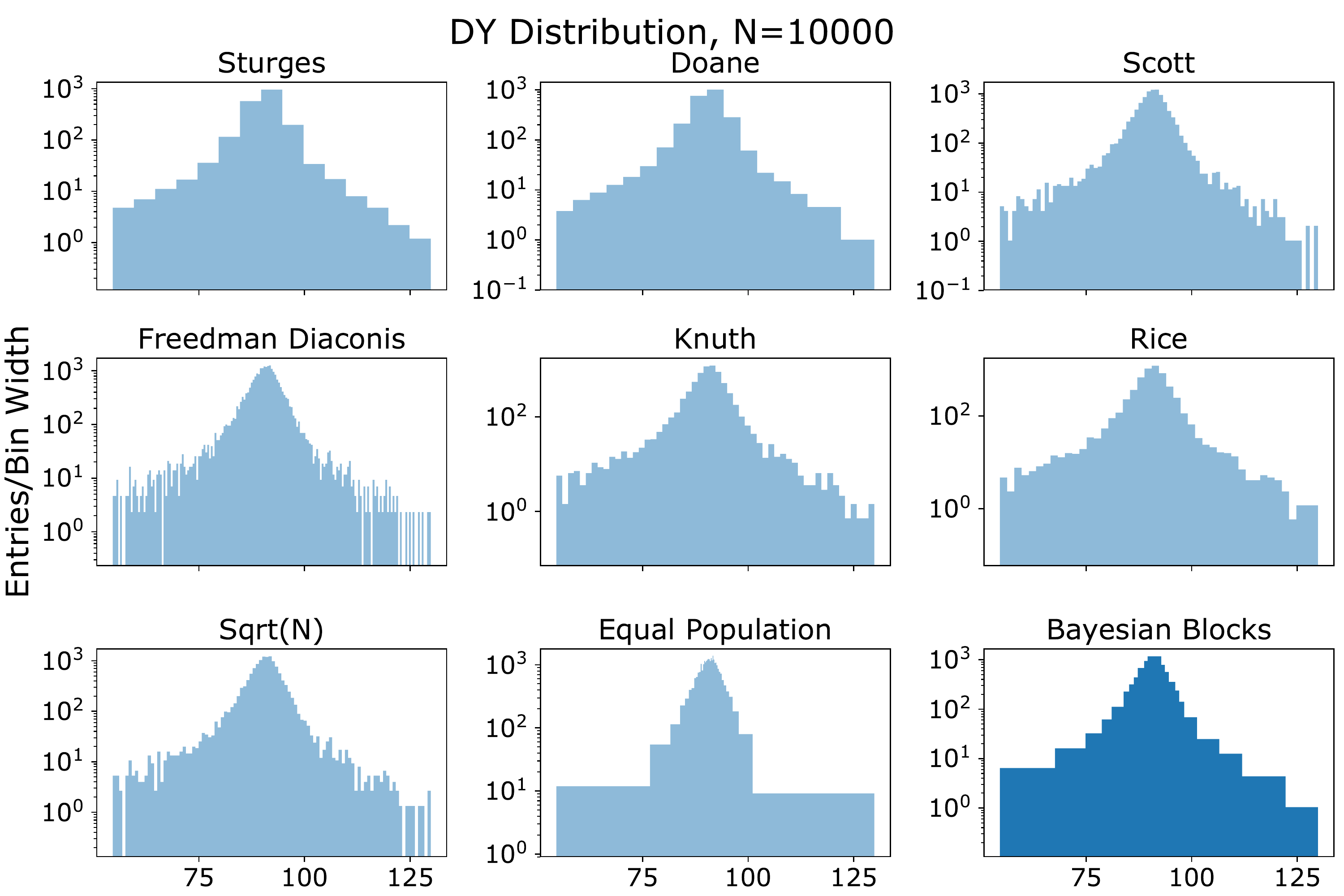}
        \vspace*{2mm}
    \end{subfigure}
    \caption{Histograms of Drell-Yan (DY) distribution for different sized datasets.\label{fig:hist_DY}}
\end{figure*}
\begin{figure*}[!htb]
    \centering
    \begin{subfigure}{\columnwidth}
        \centering
        \includegraphics[width=0.9\columnwidth]{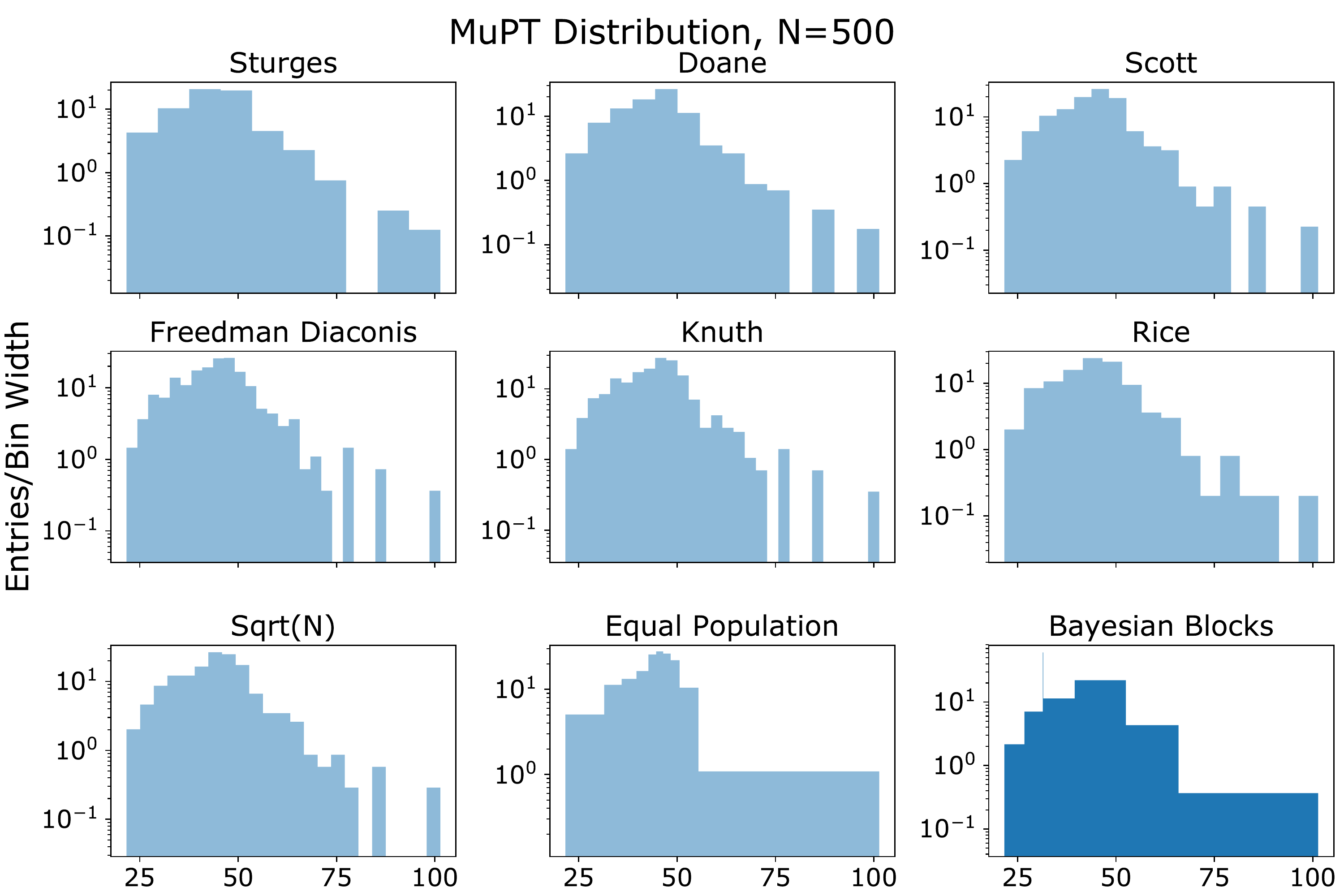}
        \vspace*{2mm}
    \end{subfigure}
    \begin{subfigure}{\columnwidth}
        \centering
        \includegraphics[width=0.9\columnwidth]{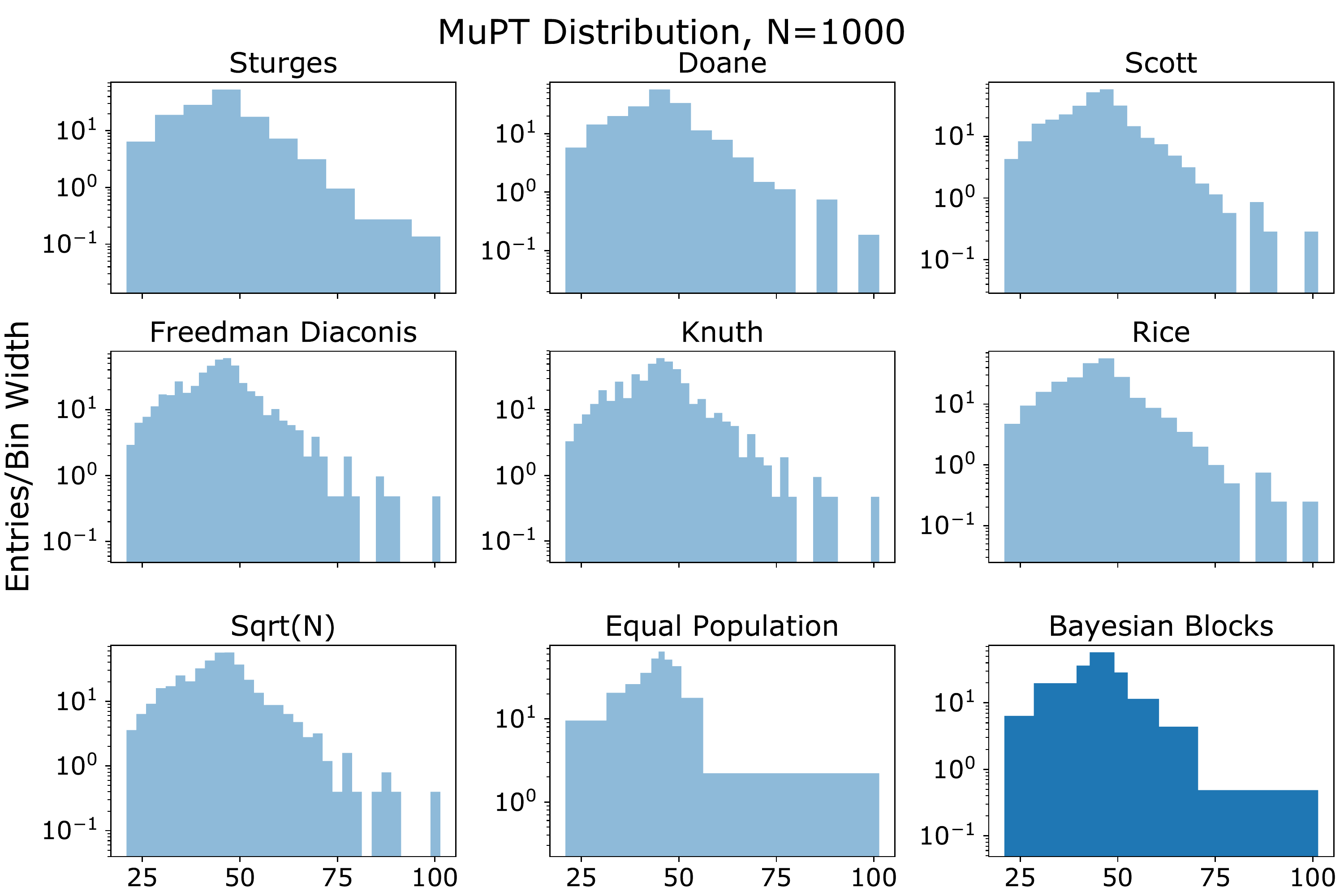}
        \vspace*{2mm}
    \end{subfigure}
    \begin{subfigure}{\columnwidth}
        \centering
        \includegraphics[width=0.9\columnwidth]{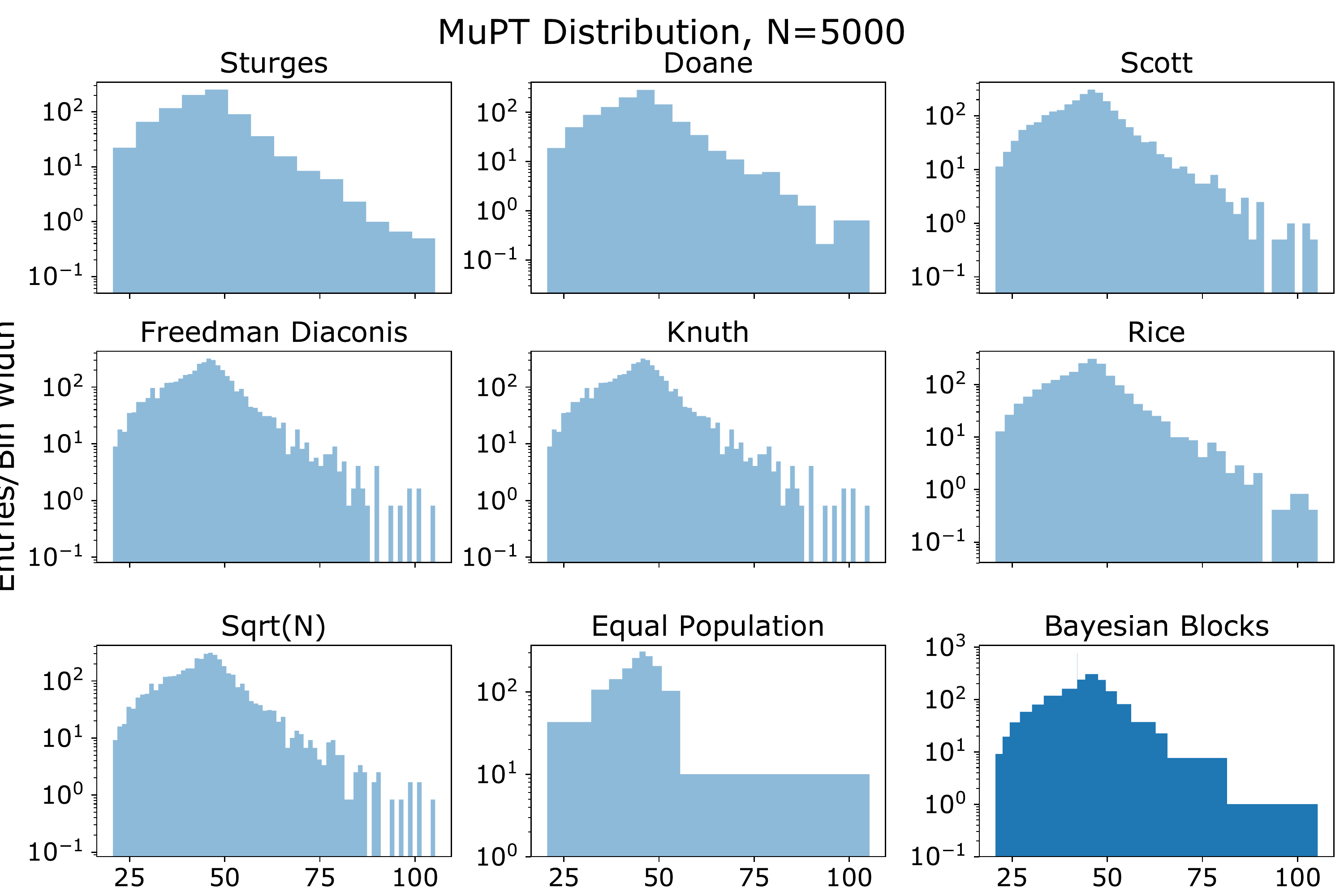}
        \vspace*{2mm}
    \end{subfigure}
    \begin{subfigure}{\columnwidth}
        \centering
        \includegraphics[width=0.9\columnwidth]{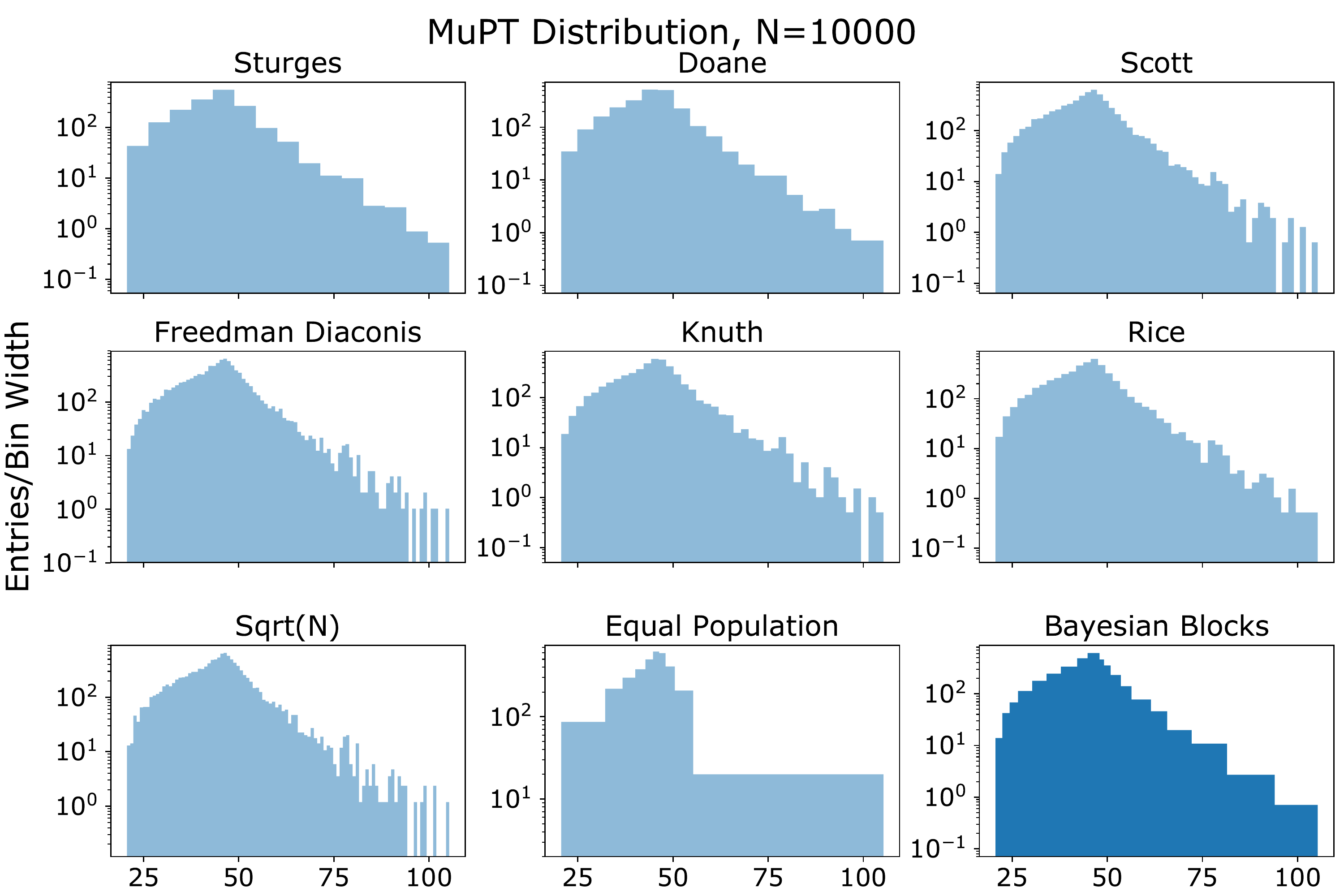}
        \vspace*{2mm}
    \end{subfigure}
    \caption{Histograms of muon transverse momentum (MuPT) for different sized datasets.\label{fig:hist_MuPT}}
\end{figure*}
\begin{figure*}[!htb]
    \centering
    \begin{subfigure}{\columnwidth}
        \centering
        \includegraphics[width=0.9\columnwidth]{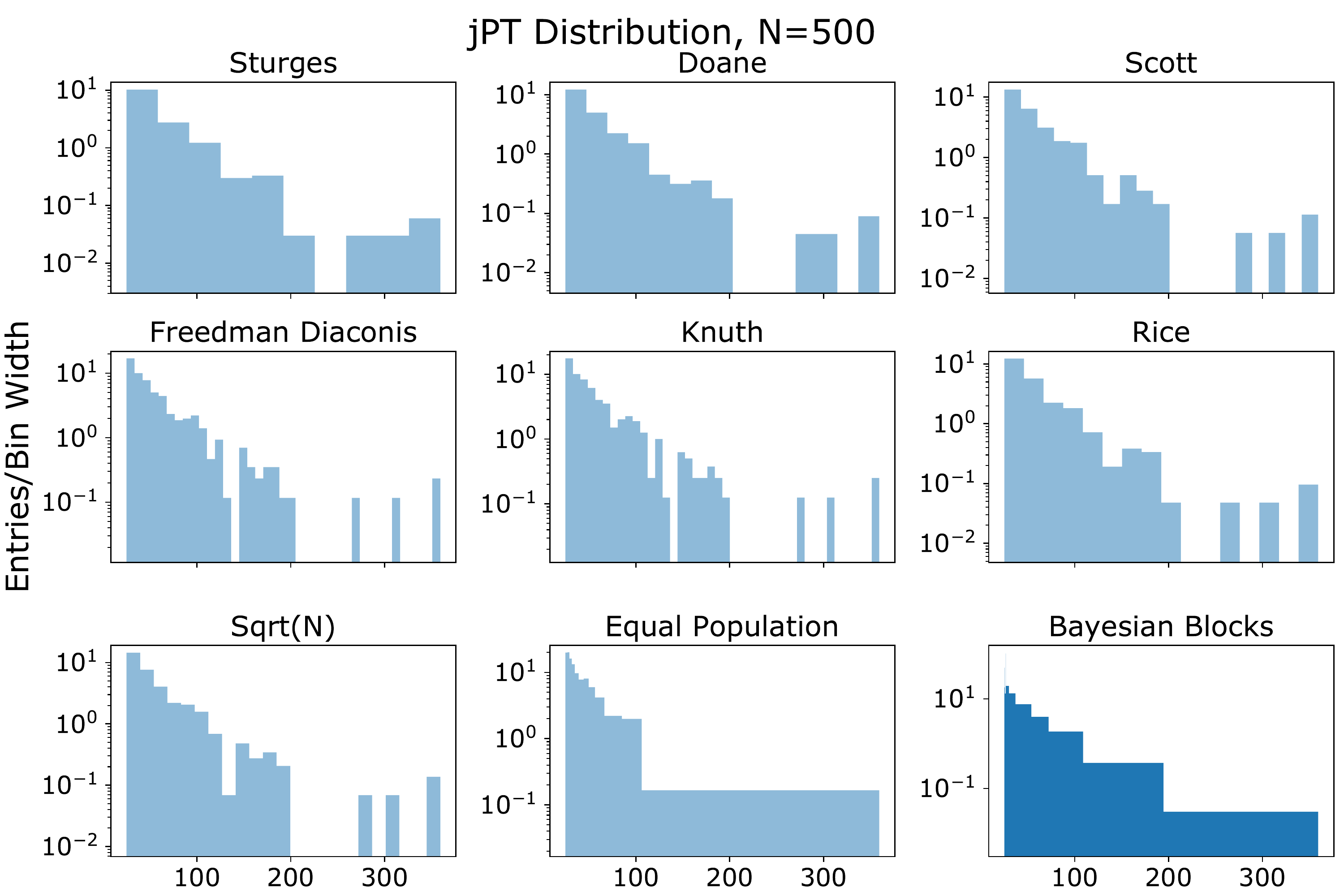}
        \vspace*{2mm}
    \end{subfigure}
    \begin{subfigure}{\columnwidth}
        \centering
        \includegraphics[width=0.9\columnwidth]{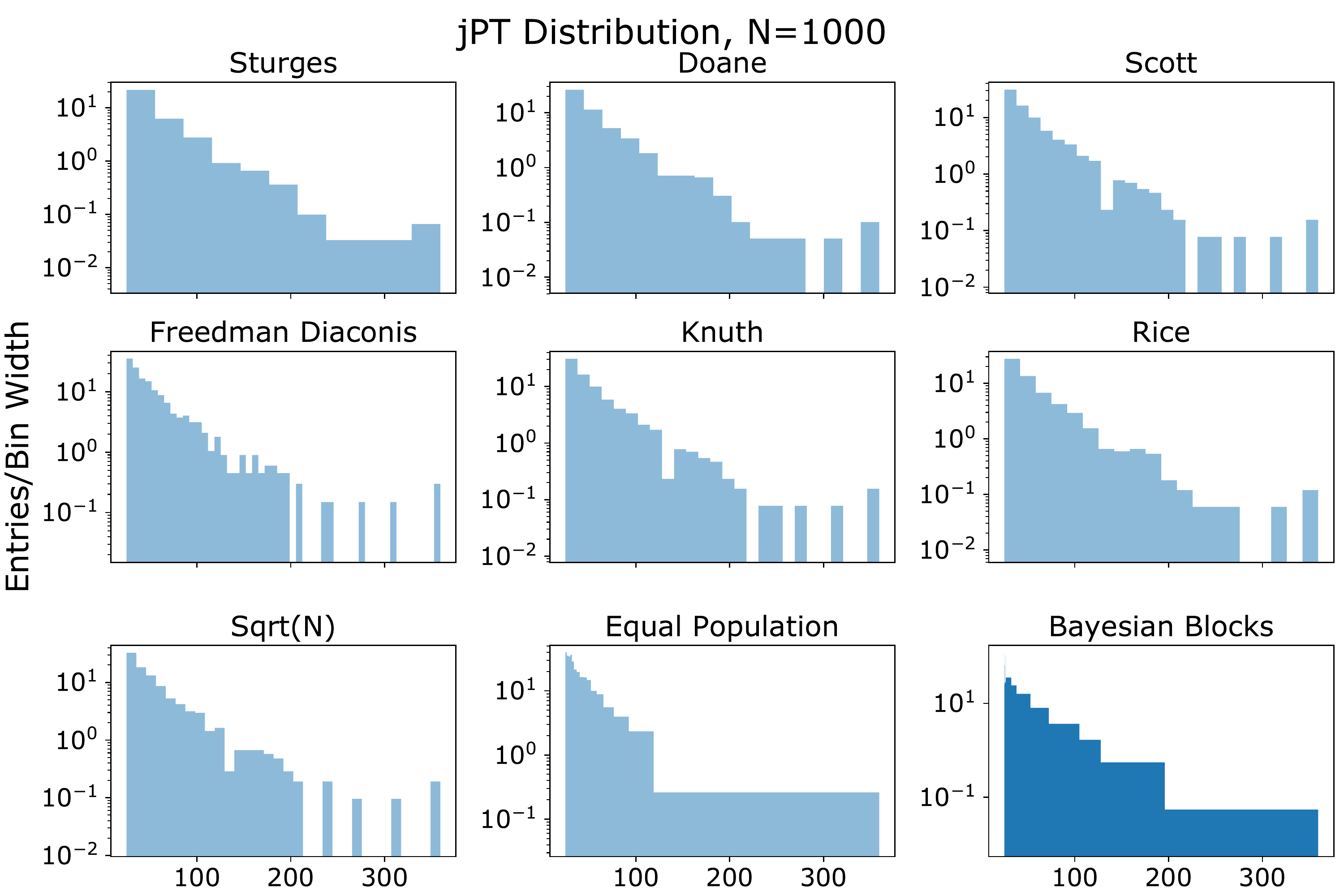}
        \vspace*{2mm}
    \end{subfigure}
    \begin{subfigure}{\columnwidth}
        \centering
        \includegraphics[width=0.9\columnwidth]{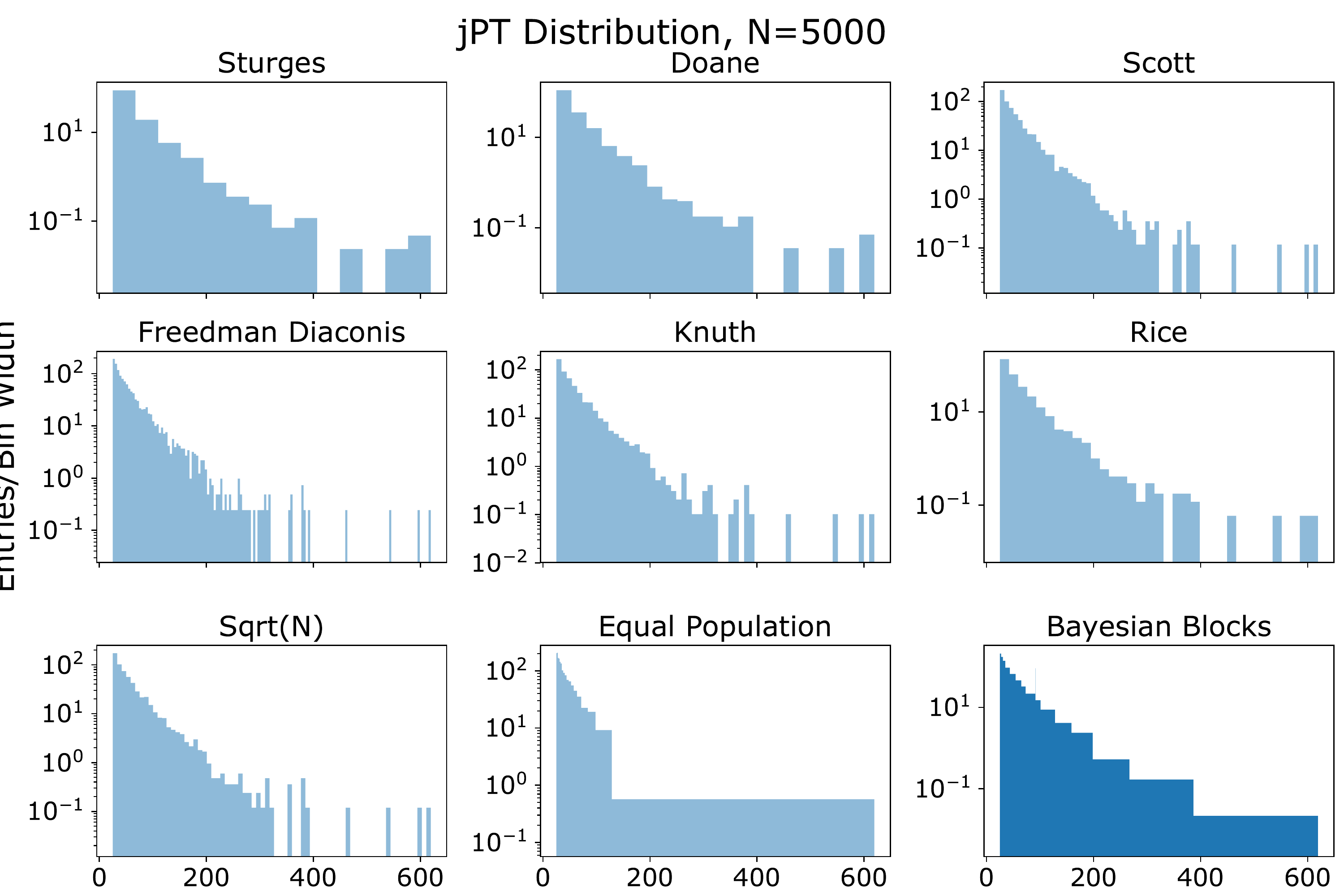}
        \vspace*{2mm}
    \end{subfigure}
    \begin{subfigure}{\columnwidth}
        \centering
        \includegraphics[width=0.9\columnwidth]{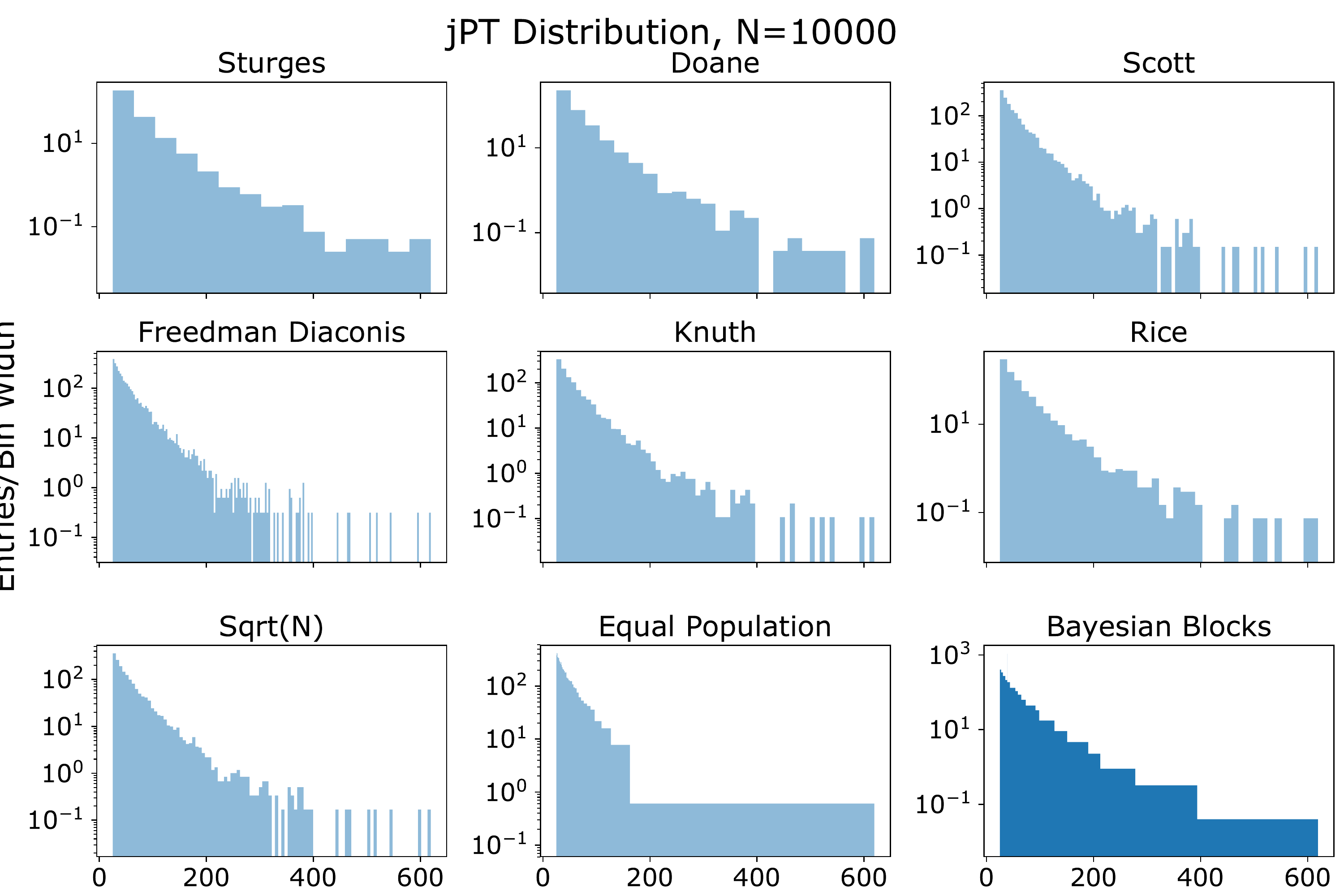}
        \vspace*{2mm}
    \end{subfigure}
    \caption{Histograms of jet transverse momentum (jPT) for different sized datasets.\label{fig:hist_jPT}}
\end{figure*}
\begin{figure*}[!htb]
    \centering
    \begin{subfigure}{\columnwidth}
        \centering
        \includegraphics[width=0.9\columnwidth]{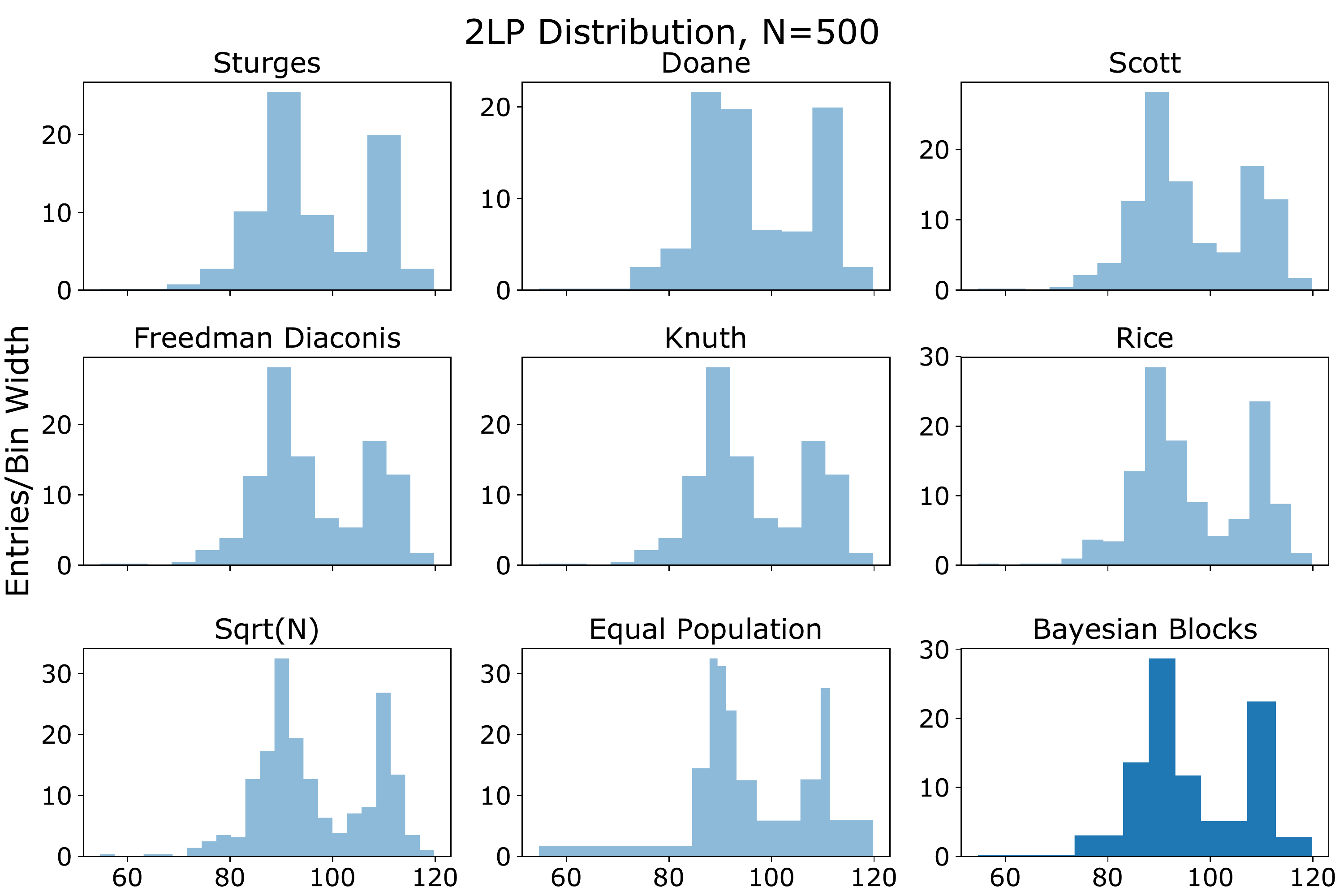}
        \vspace*{2mm}
    \end{subfigure}
    \begin{subfigure}{\columnwidth}
        \centering
        \includegraphics[width=0.9\columnwidth]{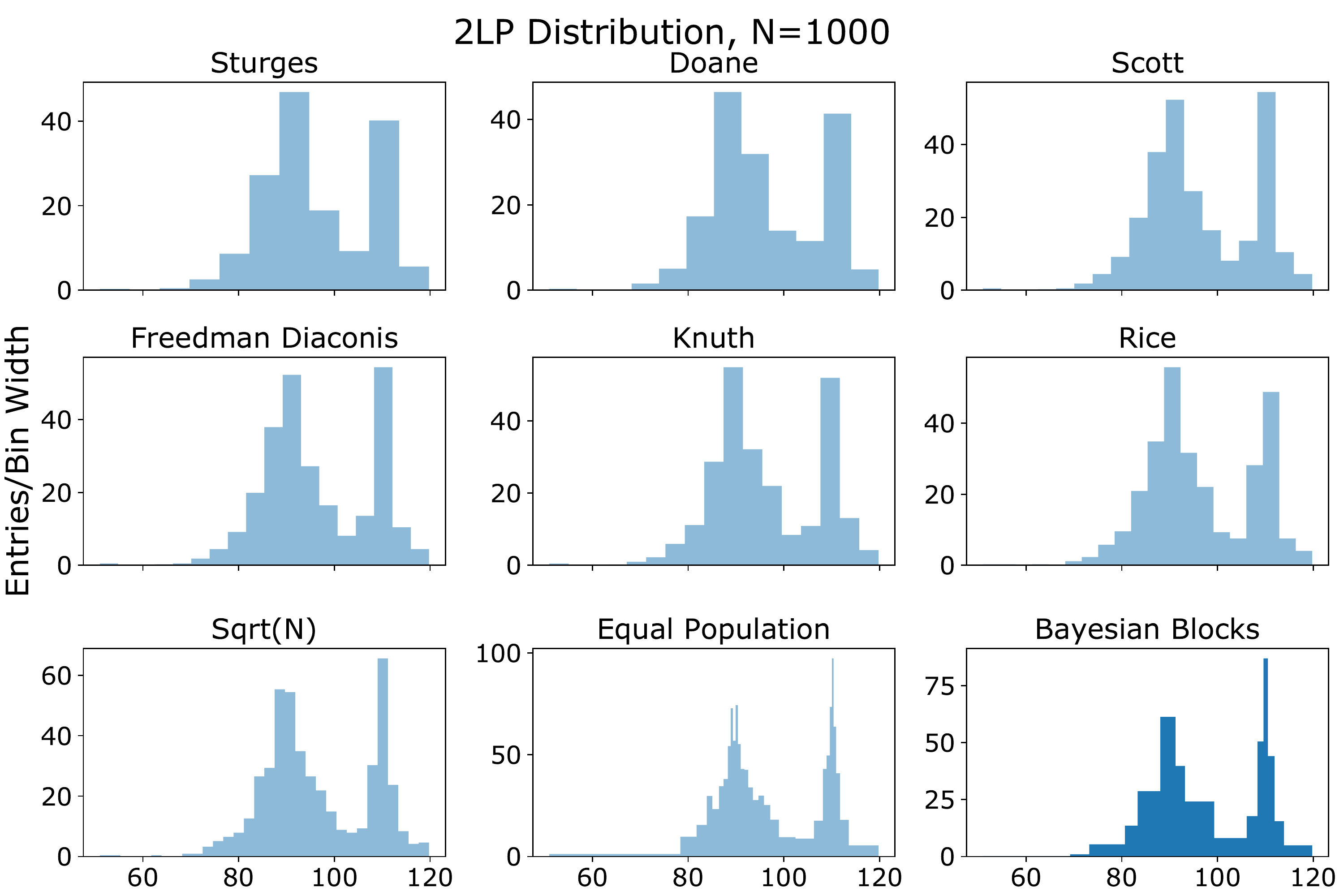}
        \vspace*{2mm}
    \end{subfigure}
    \begin{subfigure}{\columnwidth}
        \centering
        \includegraphics[width=0.9\columnwidth]{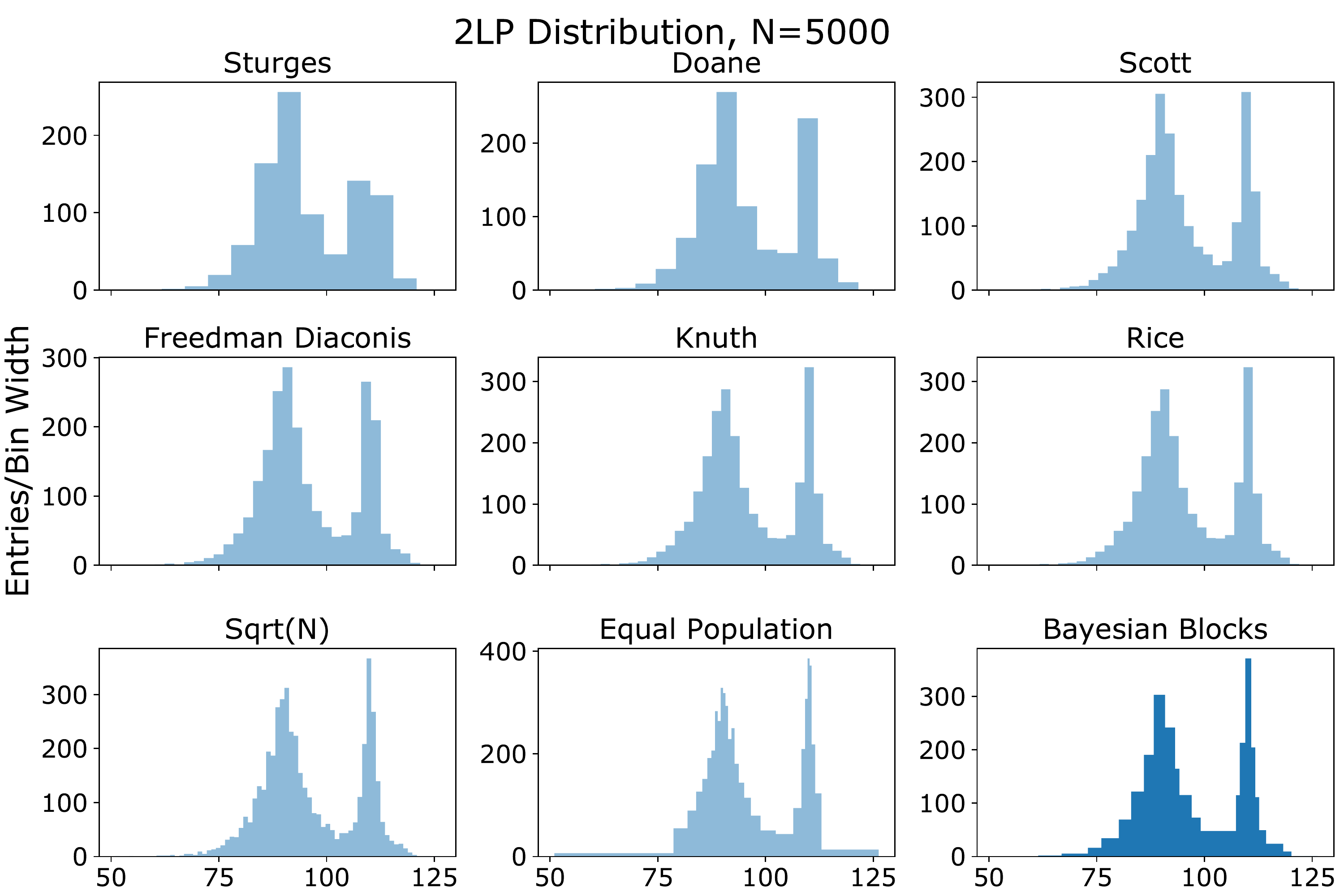}
        \vspace*{2mm}
    \end{subfigure}
    \begin{subfigure}{\columnwidth}
        \centering
        \includegraphics[width=0.9\columnwidth]{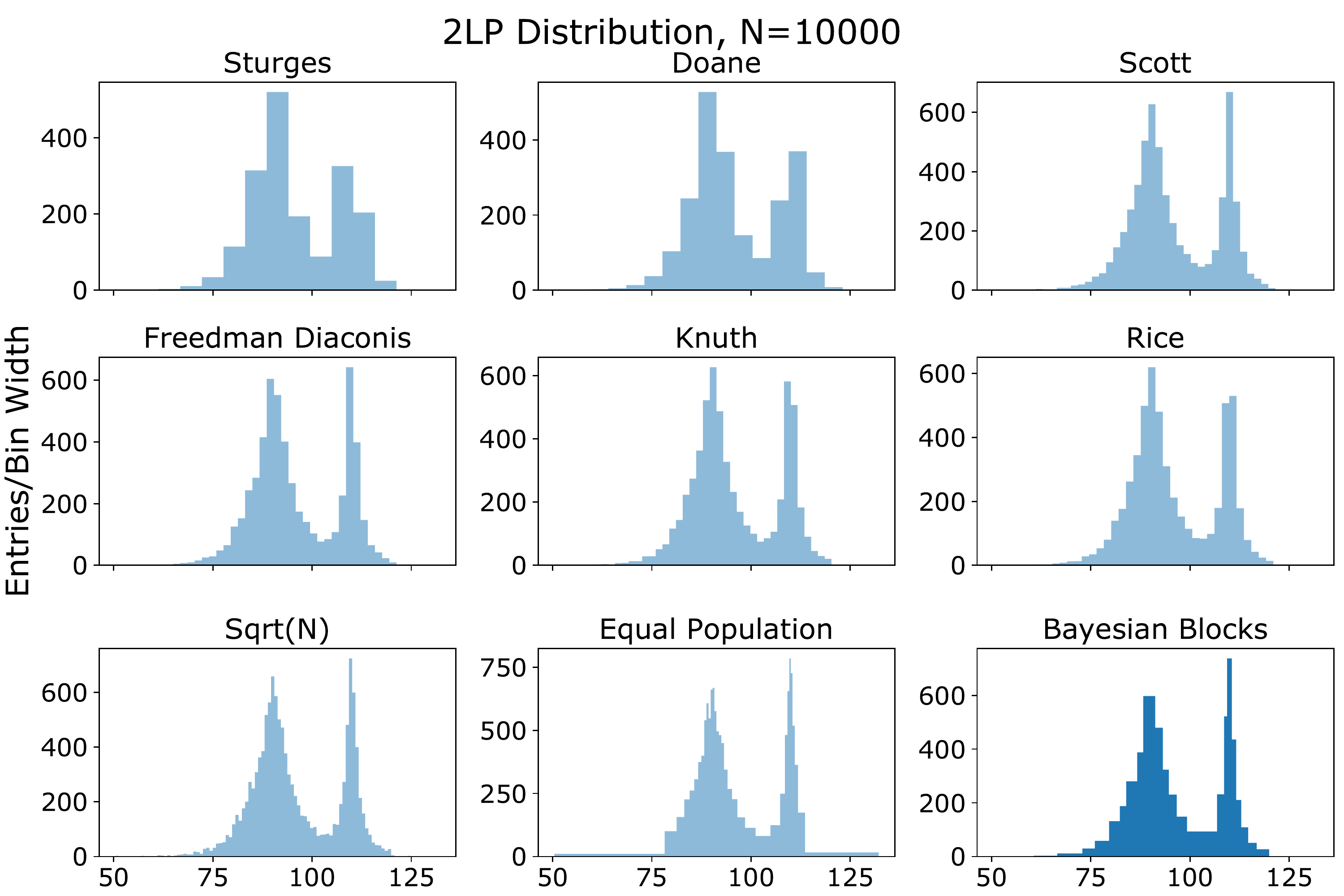}
        \vspace*{2mm}
    \end{subfigure}
    \caption{Histograms of a bimodal distribution (2LP) for different sized datasets.\label{fig:hist_2LP}}
\end{figure*}
\begin{figure*}[!htb]
    \centering
    \begin{subfigure}{\columnwidth}
        \centering
        \includegraphics[width=0.9\columnwidth]{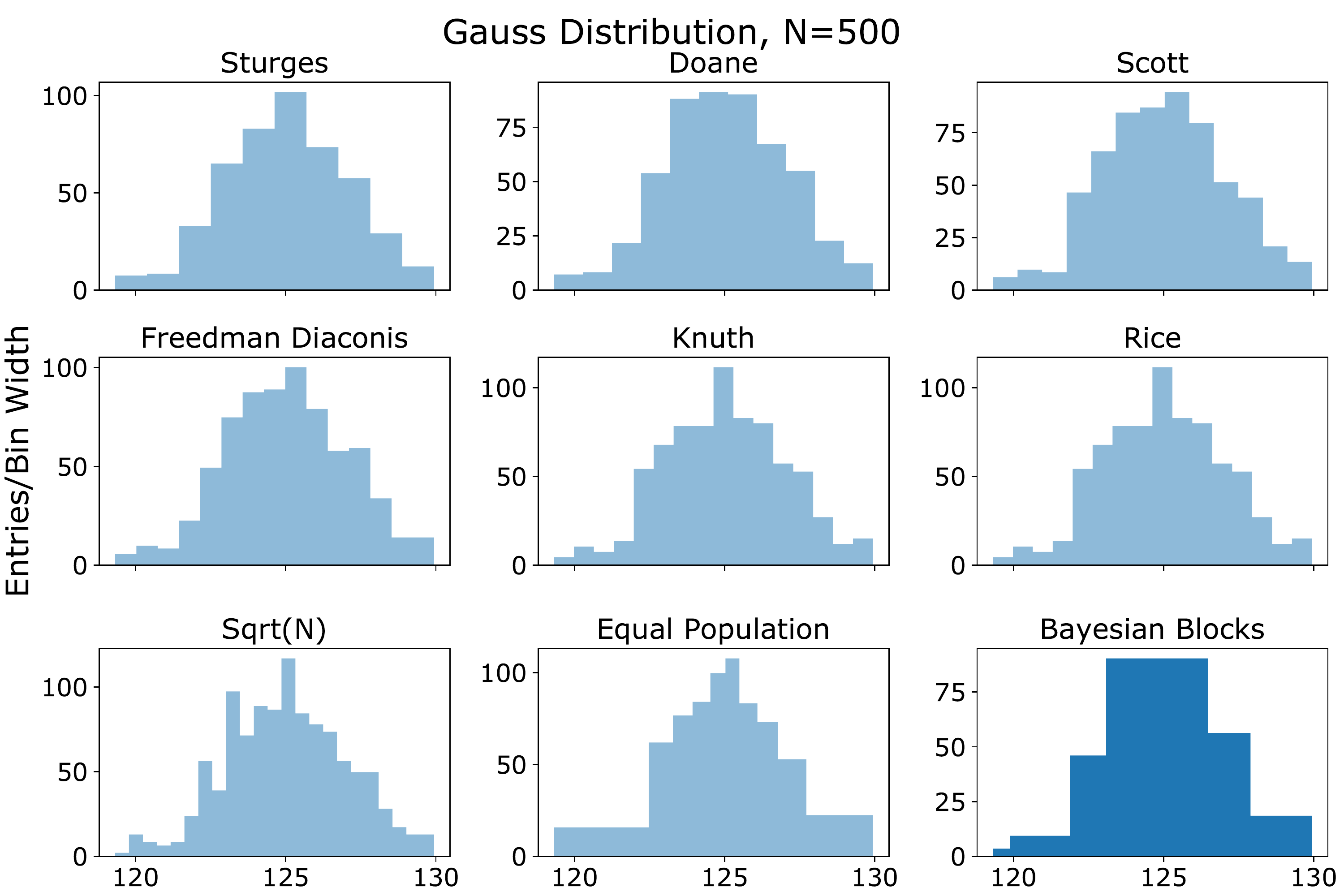}
        \vspace*{2mm}
    \end{subfigure}
    \begin{subfigure}{\columnwidth}
        \centering
        \includegraphics[width=0.9\columnwidth]{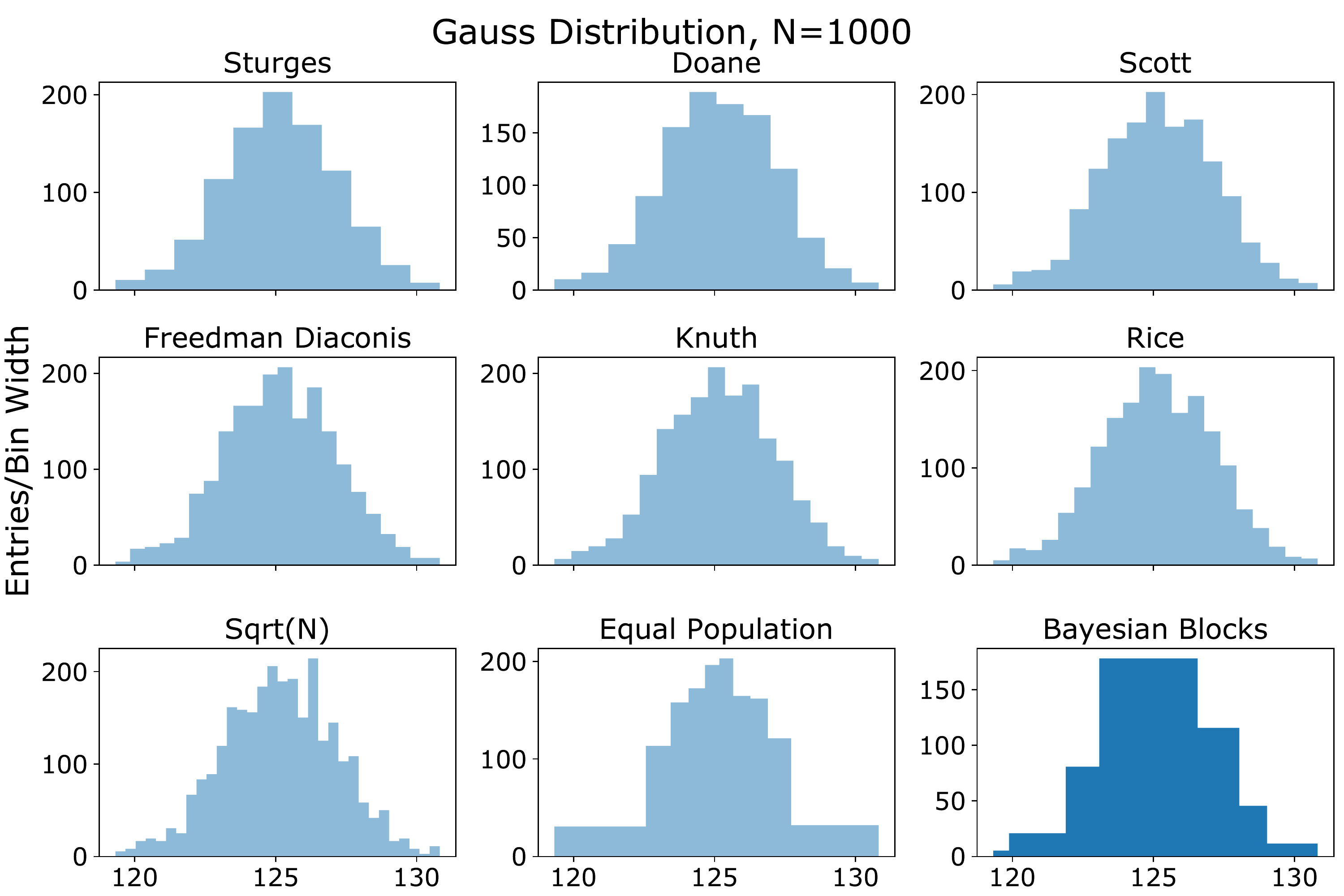}
        \vspace*{2mm}
    \end{subfigure}
    \begin{subfigure}{\columnwidth}
        \centering
        \includegraphics[width=0.9\columnwidth]{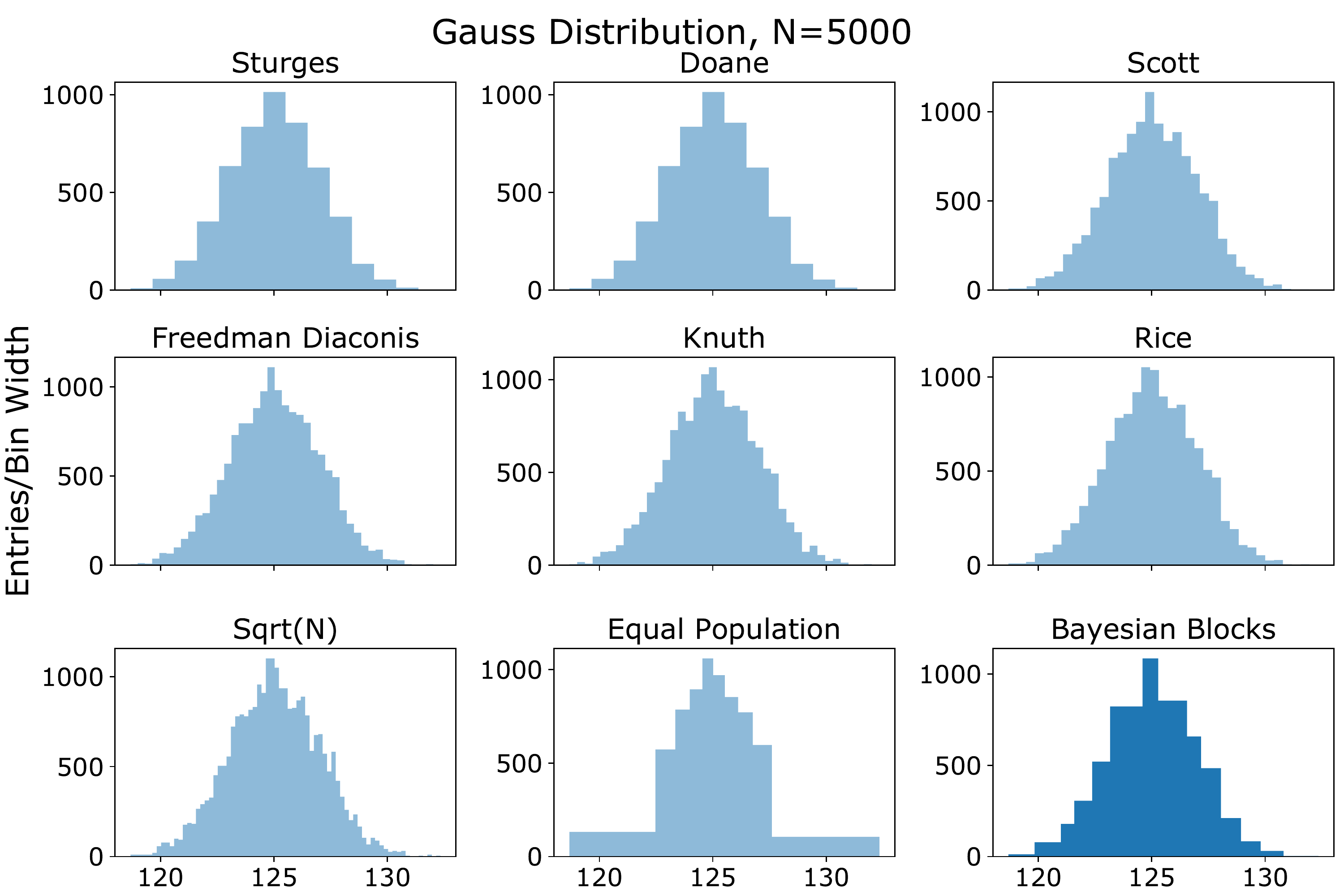}
        \vspace*{2mm}
    \end{subfigure}
    \begin{subfigure}{\columnwidth}
        \centering
        \includegraphics[width=0.9\columnwidth]{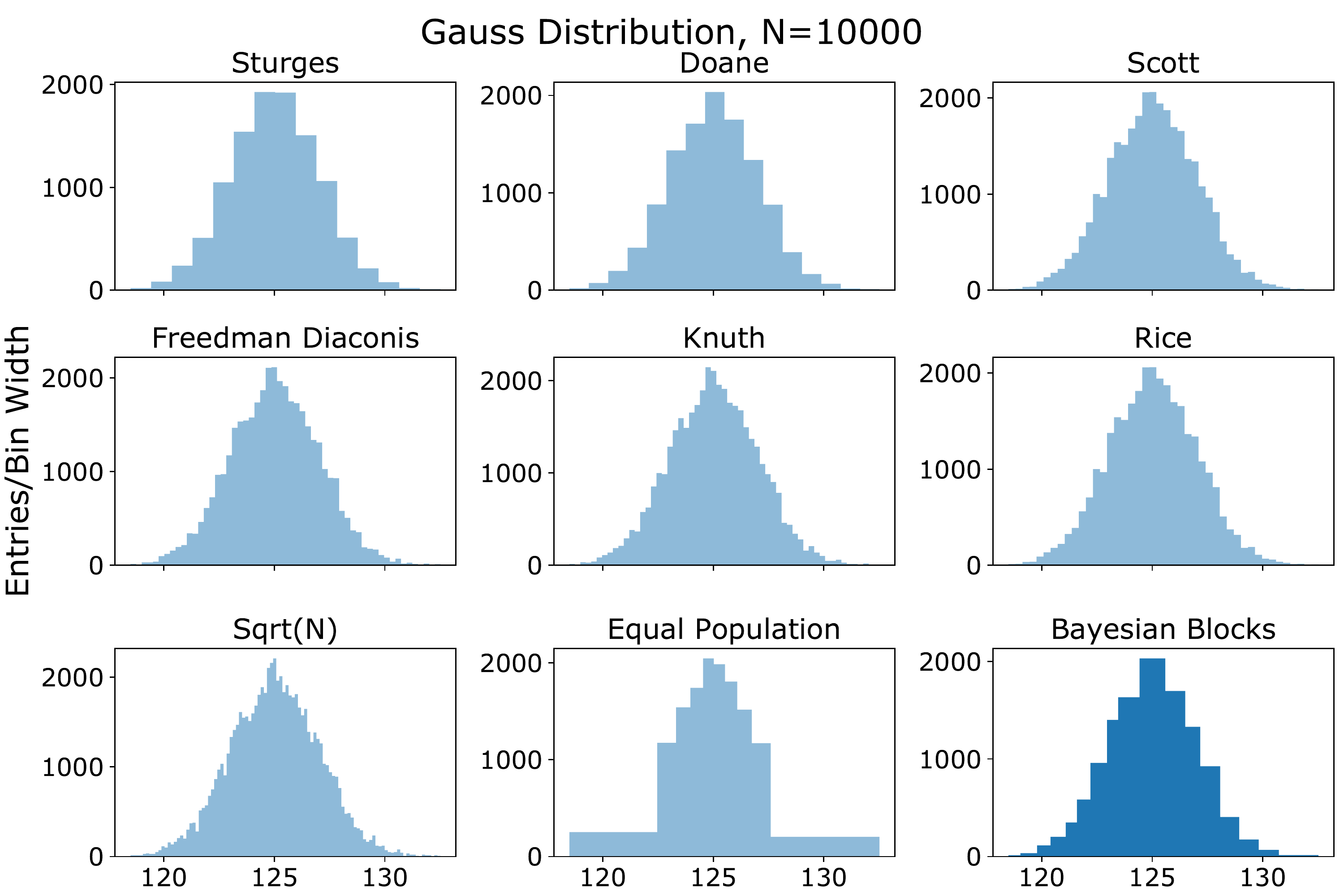}
        \vspace*{2mm}
    \end{subfigure}
    \caption{Histograms of Gaussian distribution (Gauss) for different sized datasets.\label{fig:hist_Gauss}}
\end{figure*}
\begin{figure*}[!htb]
    \centering
    \begin{subfigure}{\columnwidth}
        \centering
        \includegraphics[width=0.9\columnwidth]{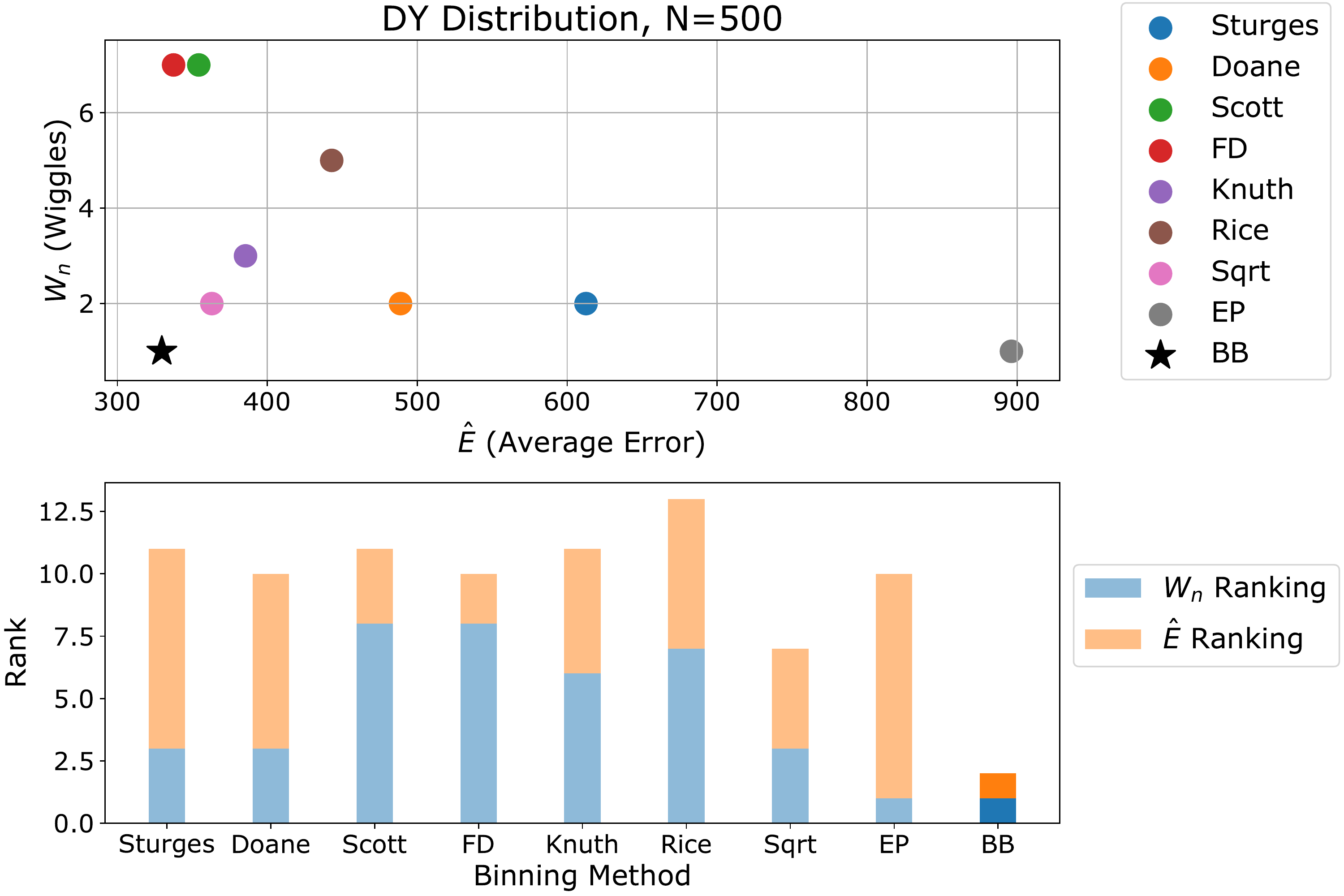}
        \vspace*{2mm}
    \end{subfigure}
    \begin{subfigure}{\columnwidth}
        \centering
        \includegraphics[width=0.9\columnwidth]{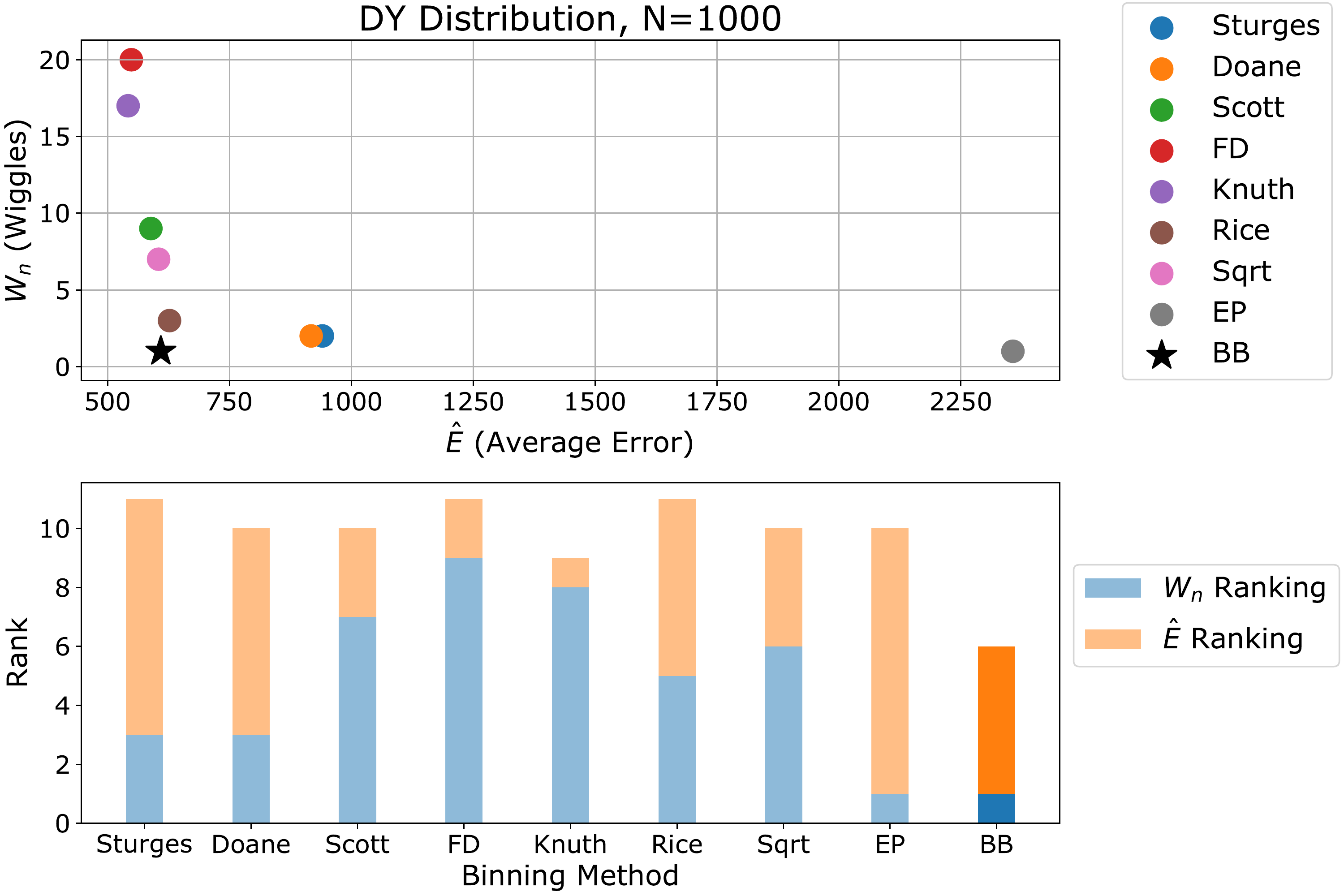}
        \vspace*{2mm}
    \end{subfigure}
    \begin{subfigure}{\columnwidth}
        \centering
        \includegraphics[width=0.9\columnwidth]{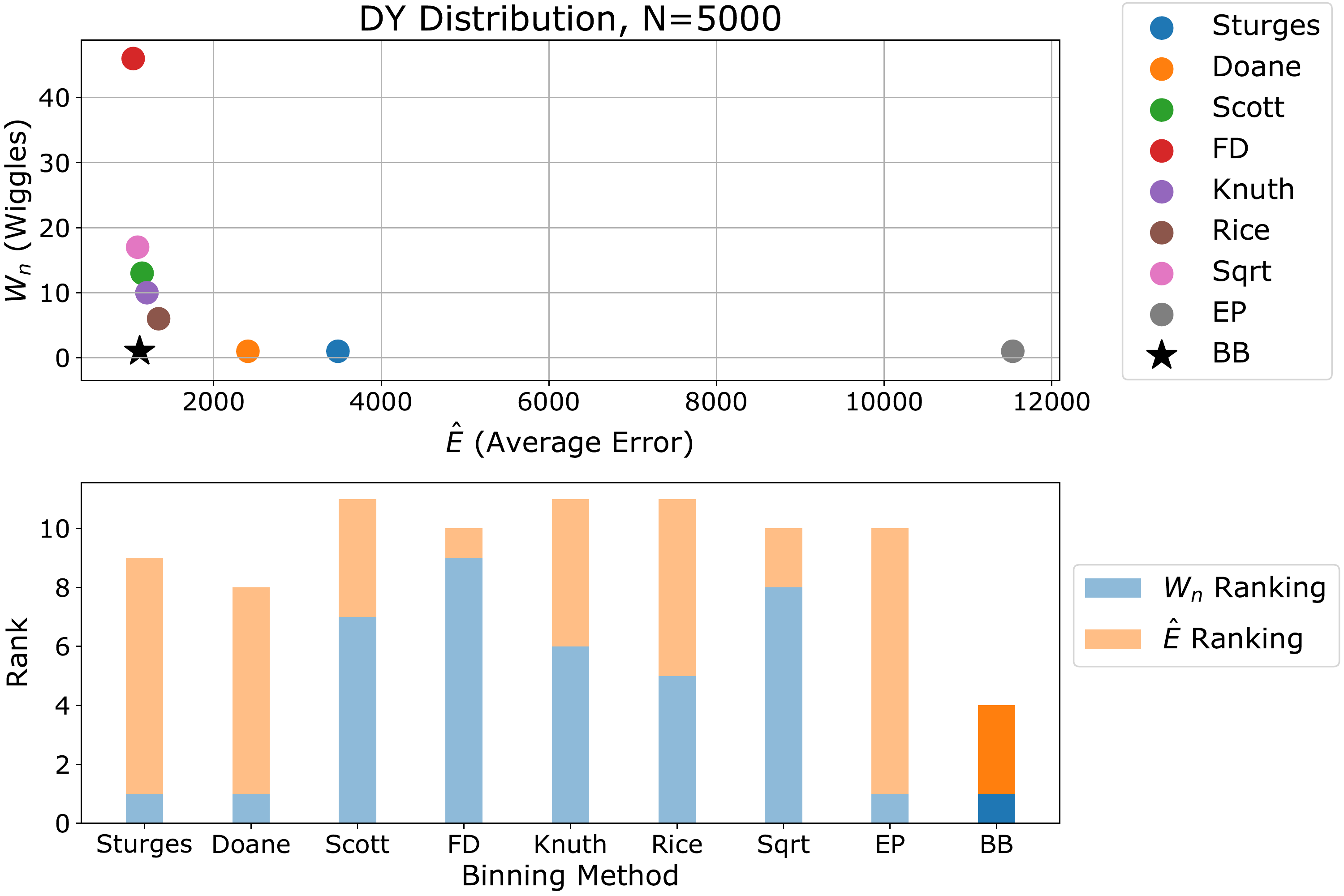}
        \vspace*{2mm}
    \end{subfigure}
    \begin{subfigure}{\columnwidth}
        \centering
        \includegraphics[width=0.9\columnwidth]{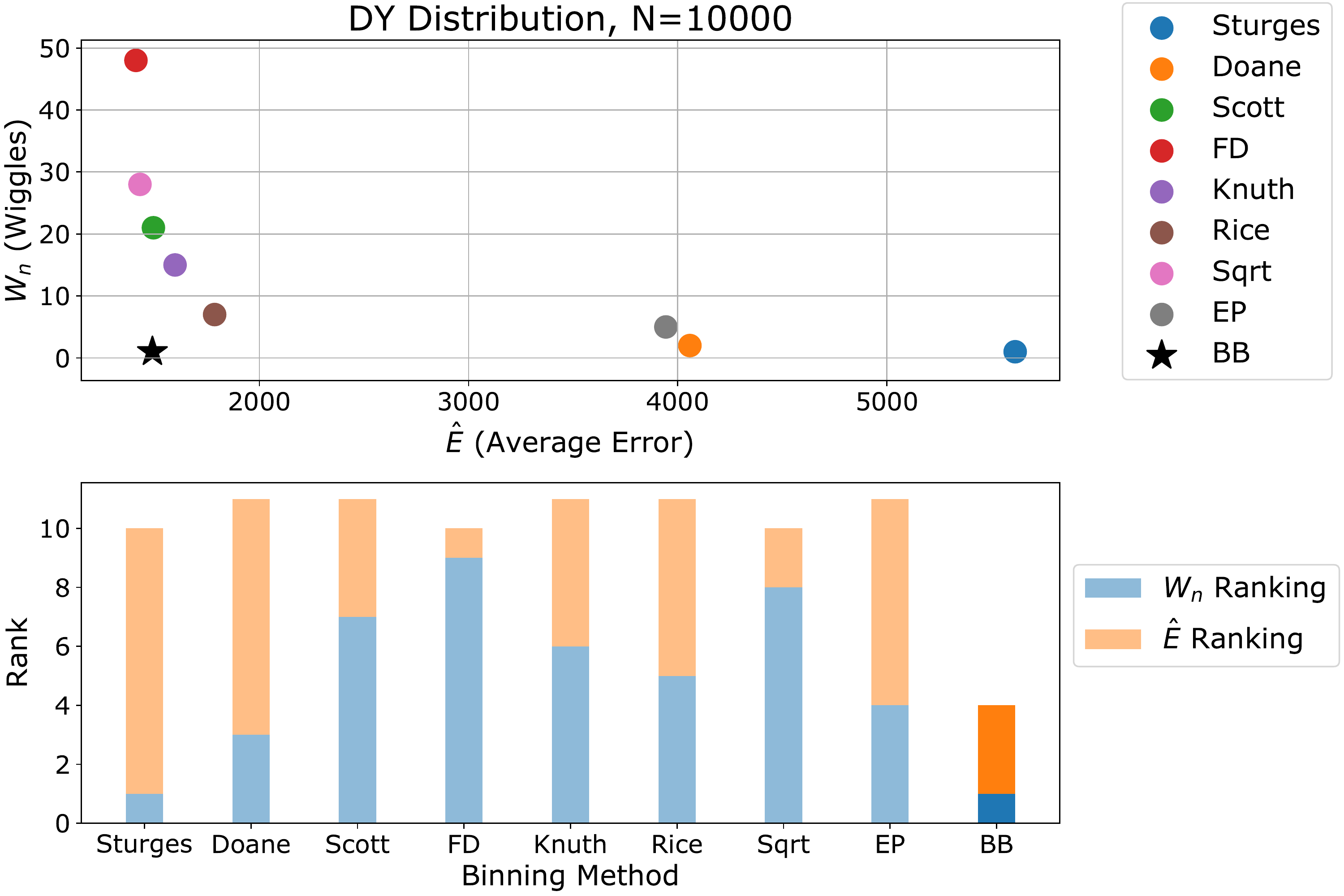}
        \vspace*{2mm}
    \end{subfigure}
    \caption{Metric values and combined ranks for Drell-Yan (DY) distribution for different sized datasets.\label{fig:metric_DY}}
\end{figure*}
\begin{figure*}[!htb]
    \centering
    \begin{subfigure}{\columnwidth}
        \centering
        \includegraphics[width=0.9\columnwidth]{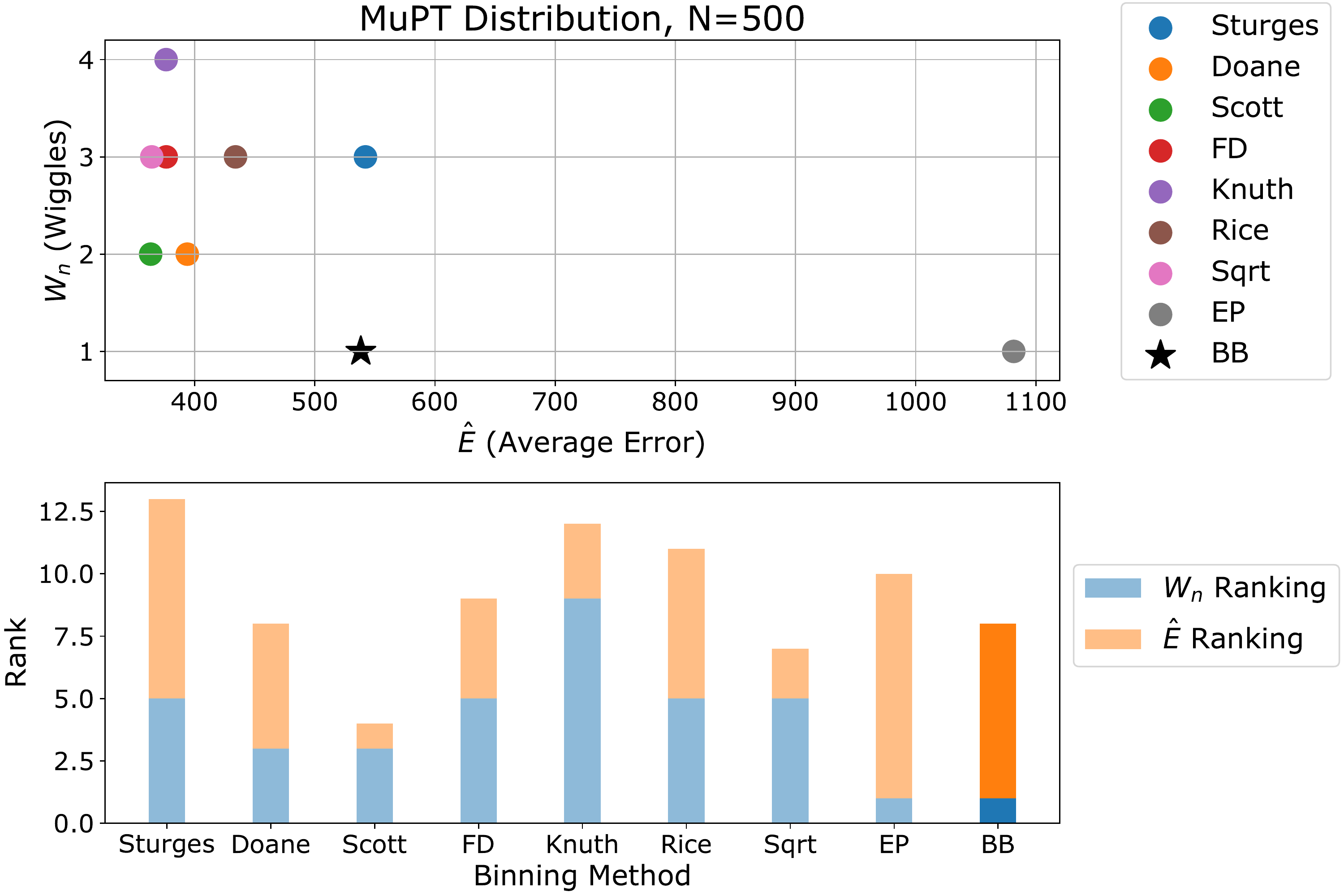}
        \vspace*{2mm}
    \end{subfigure}
    \begin{subfigure}{\columnwidth}
        \centering
        \includegraphics[width=0.9\columnwidth]{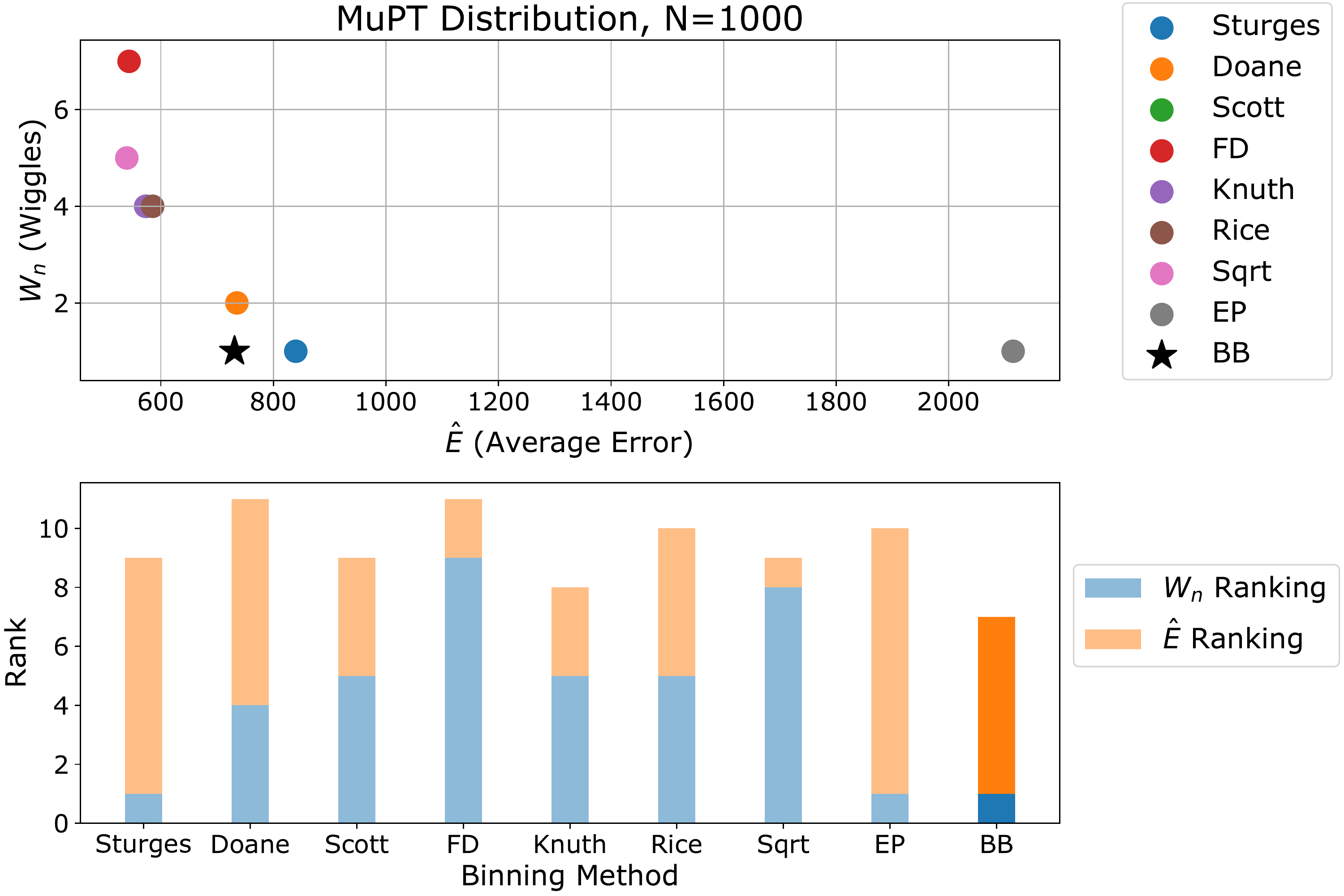}
        \vspace*{2mm}
    \end{subfigure}
    \begin{subfigure}{\columnwidth}
        \centering
        \includegraphics[width=0.9\columnwidth]{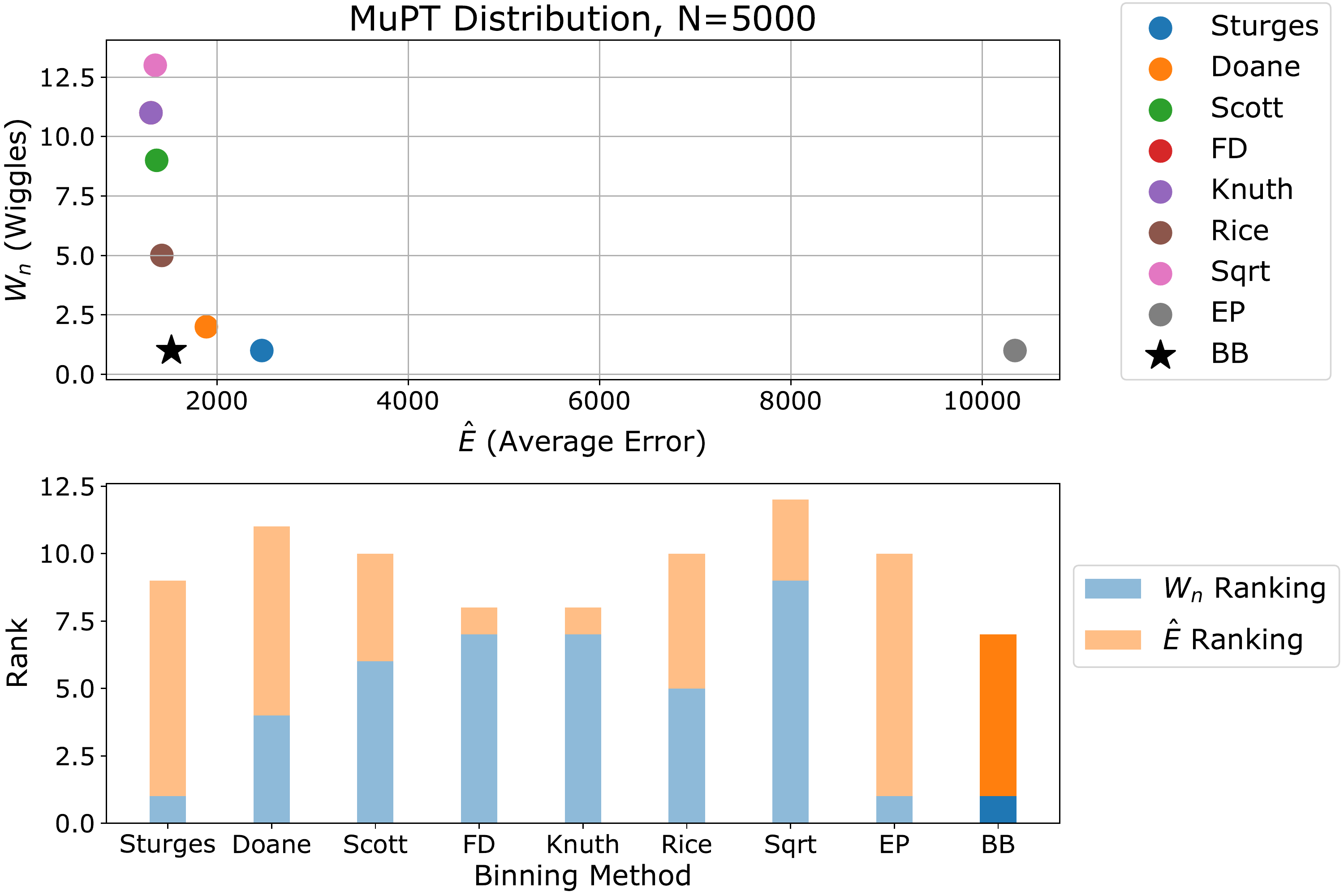}
        \vspace*{2mm}
    \end{subfigure}
    \begin{subfigure}{\columnwidth}
        \centering
        \includegraphics[width=0.9\columnwidth]{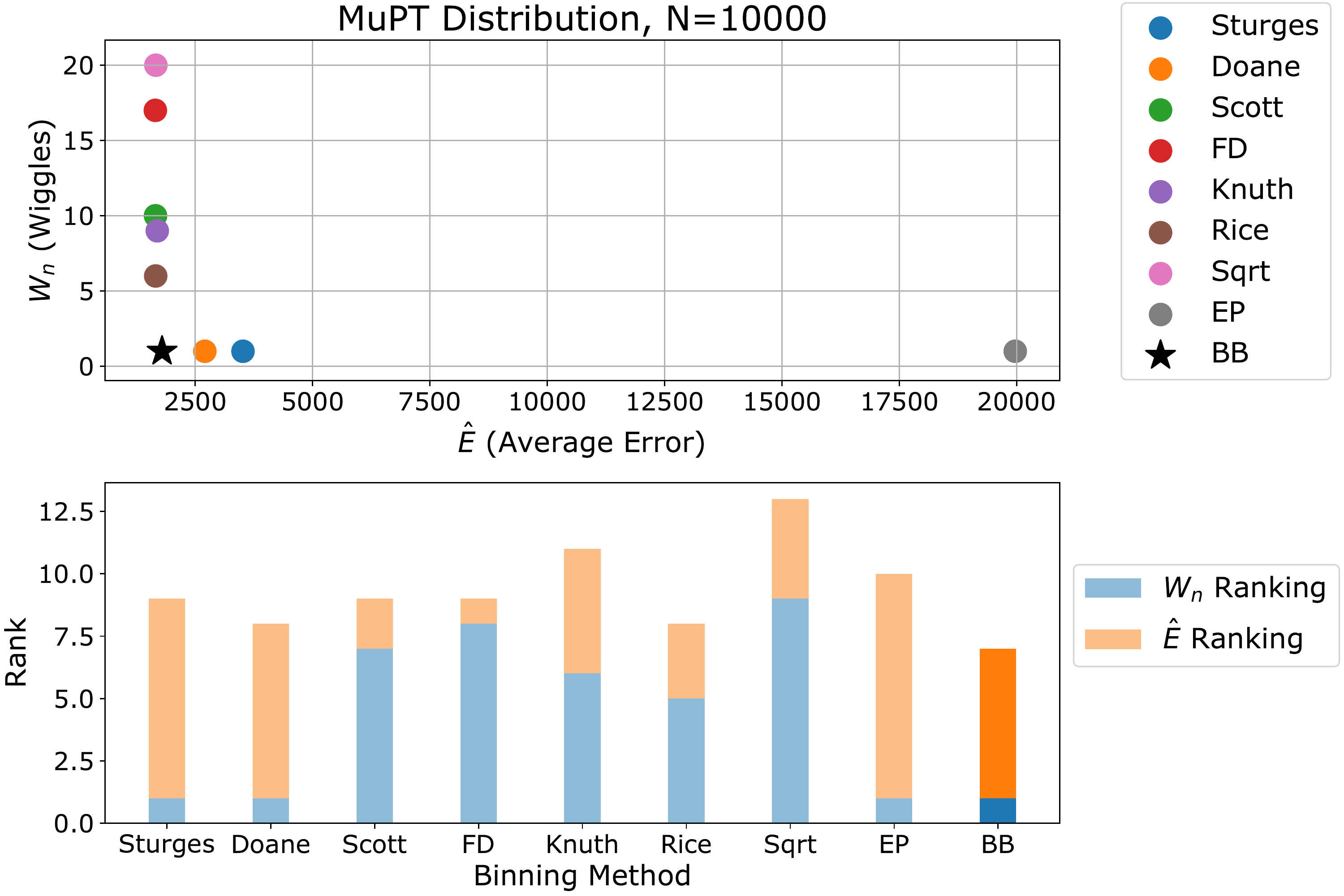}
        \vspace*{2mm}
    \end{subfigure}
    \caption{Metric values and combined ranks for muon transverse momentum (MuPT) for different sized datasets.\label{fig:metric_MuPT}}
\end{figure*}
\begin{figure*}[!htb]
    \centering
    \begin{subfigure}{\columnwidth}
        \centering
        \includegraphics[width=0.9\columnwidth]{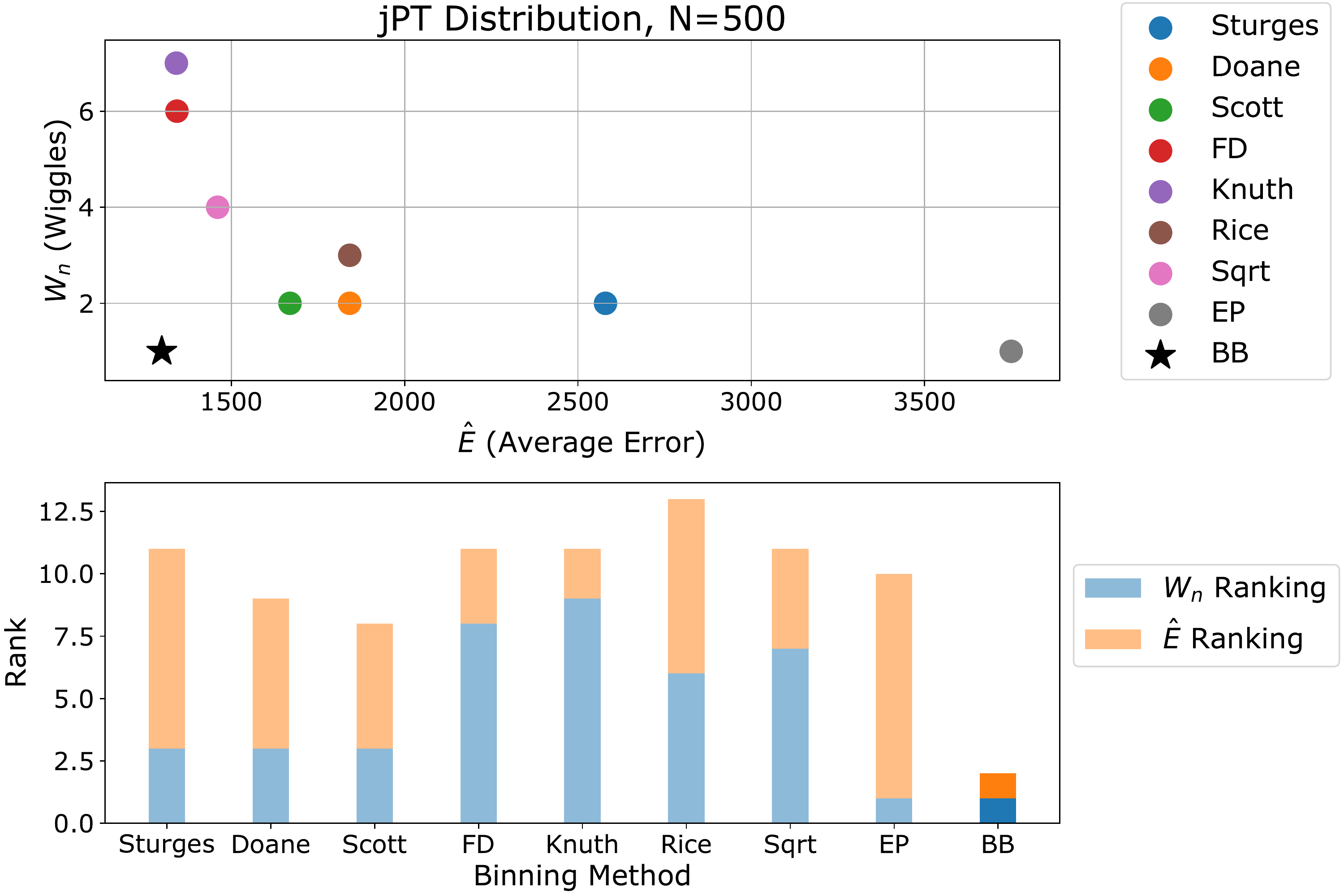}
        \vspace*{2mm}
    \end{subfigure}
    \begin{subfigure}{\columnwidth}
        \centering
        \includegraphics[width=0.9\columnwidth]{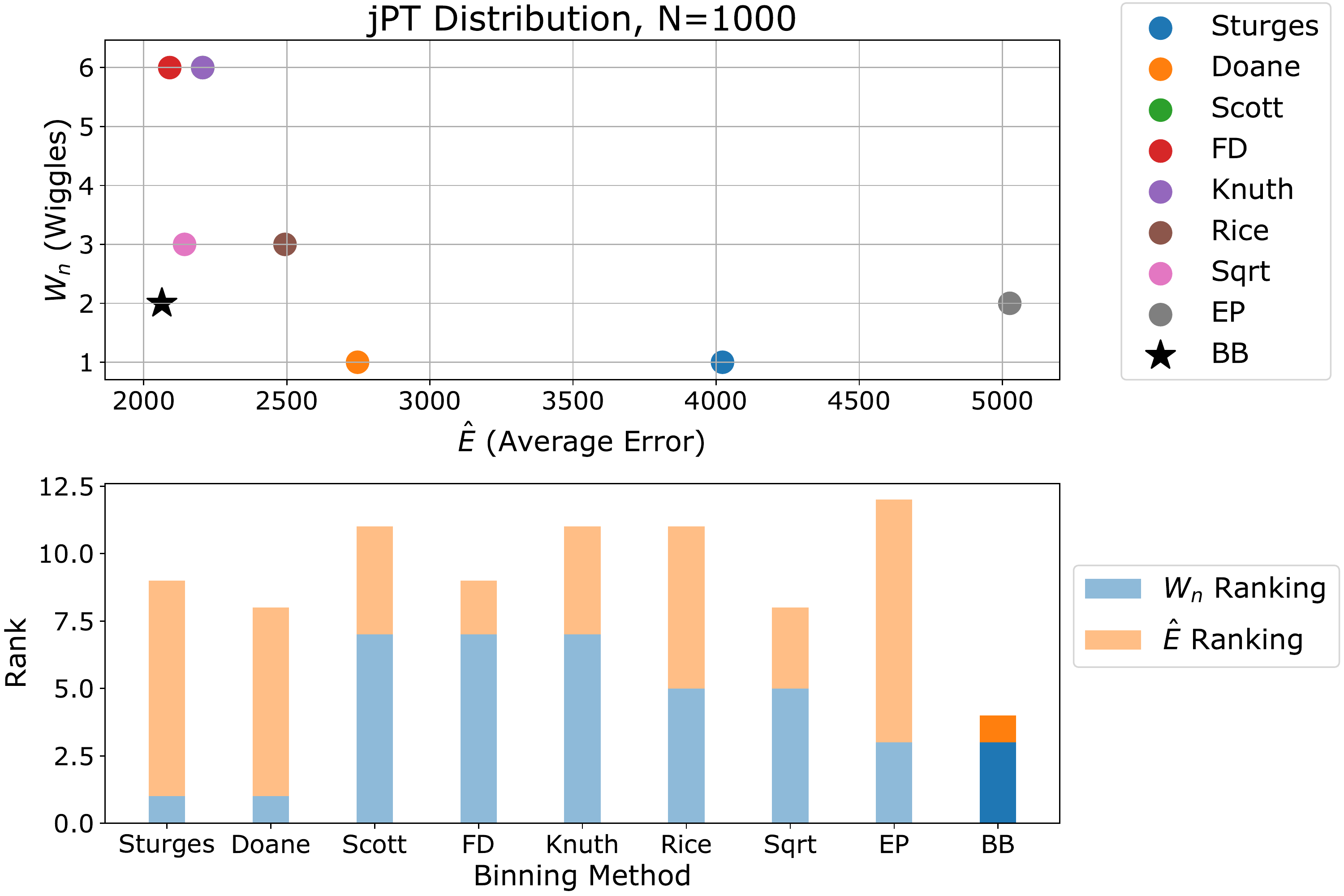}
        \vspace*{2mm}
    \end{subfigure}
    \begin{subfigure}{\columnwidth}
        \centering
        \includegraphics[width=0.9\columnwidth]{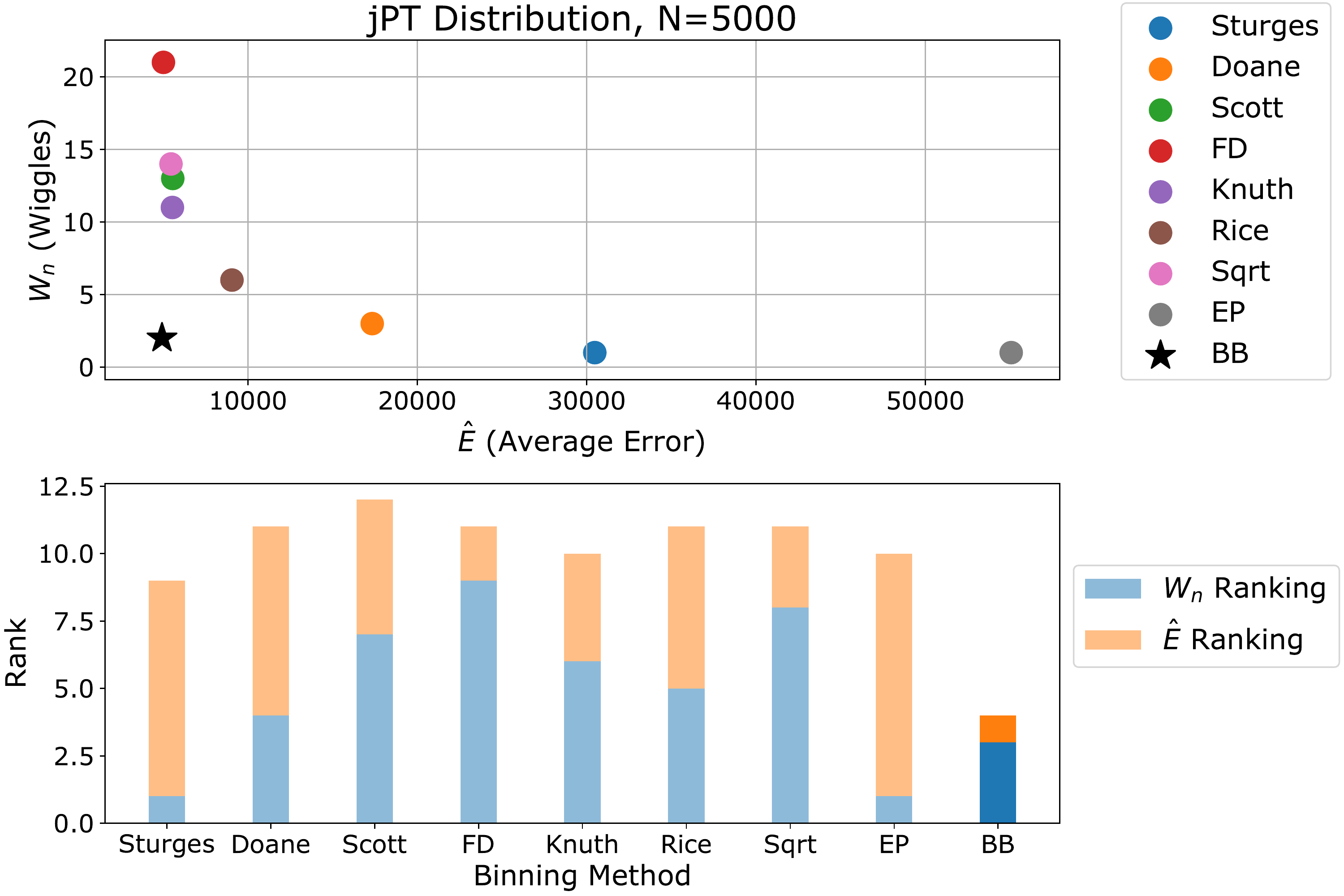}
        \vspace*{2mm}
    \end{subfigure}
    \begin{subfigure}{\columnwidth}
        \centering
        \includegraphics[width=0.9\columnwidth]{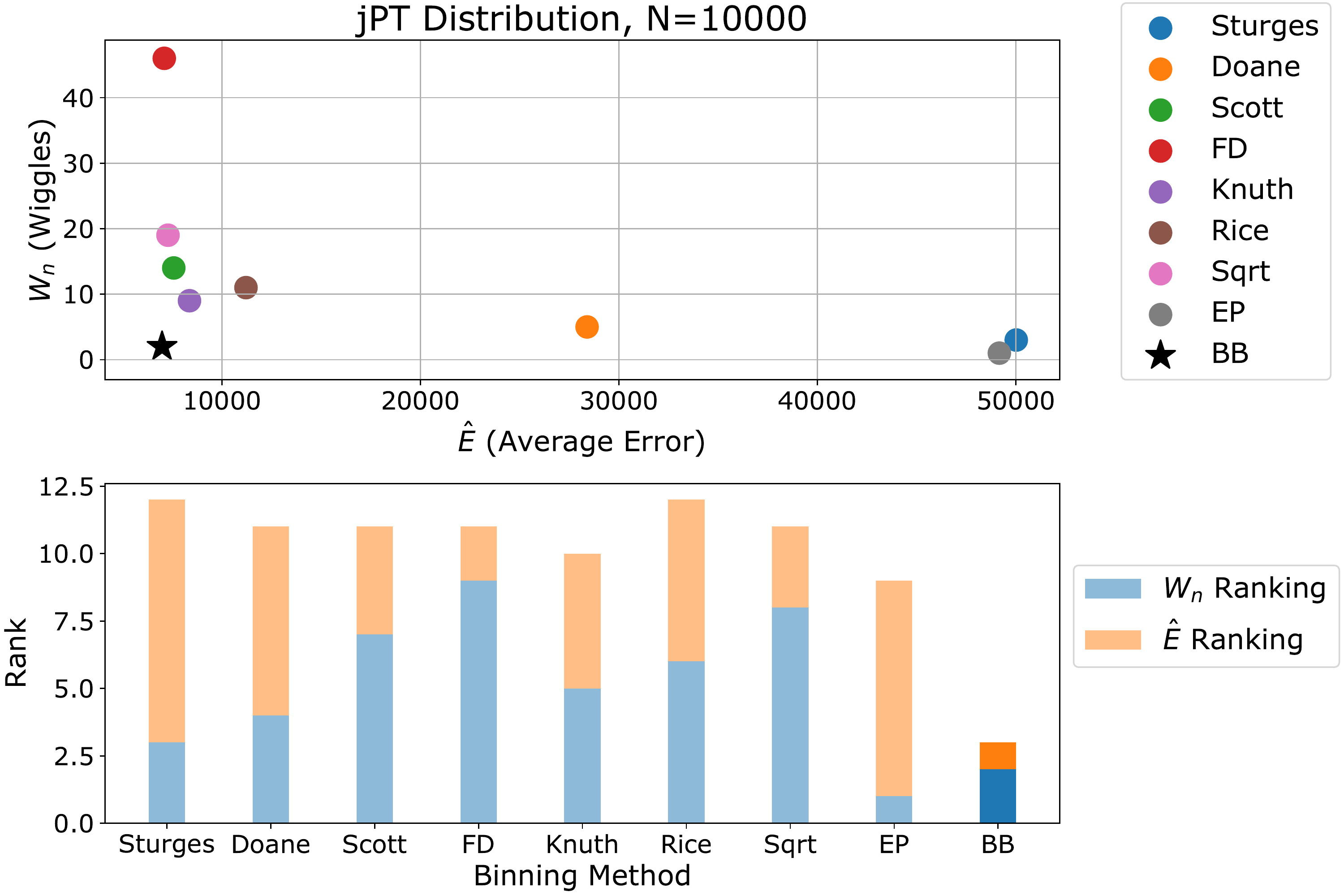}
        \vspace*{2mm}
    \end{subfigure}
    \caption{Metric values and combined ranks for jet transverse momentum (jPT) for different sized datasets.\label{fig:metric_jPT}}
\end{figure*}
\begin{figure*}[!htb]
    \centering
    \begin{subfigure}{\columnwidth}
        \centering
        \includegraphics[width=0.9\columnwidth]{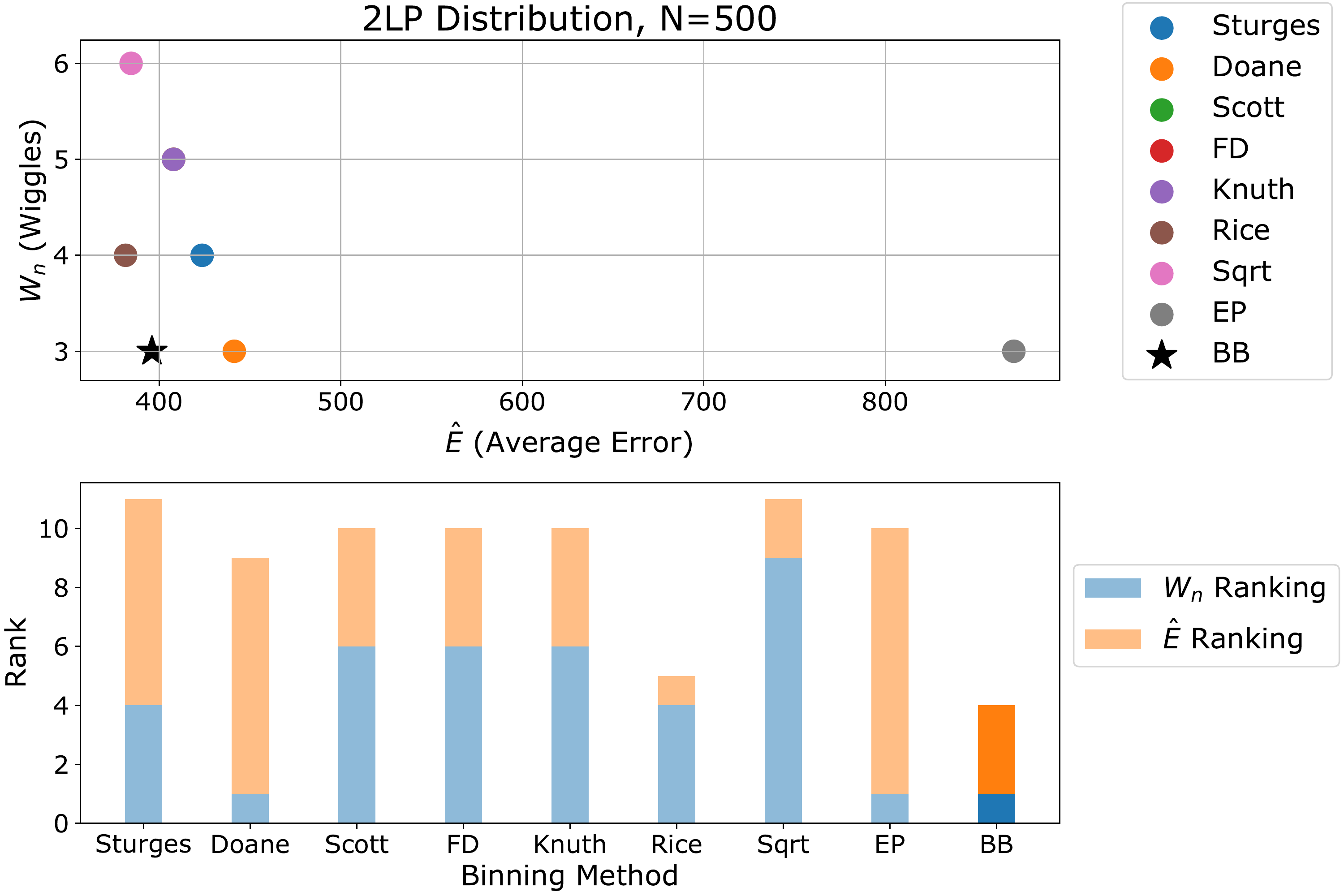}
        \vspace*{2mm}
    \end{subfigure}
    \begin{subfigure}{\columnwidth}
        \centering
        \includegraphics[width=0.9\columnwidth]{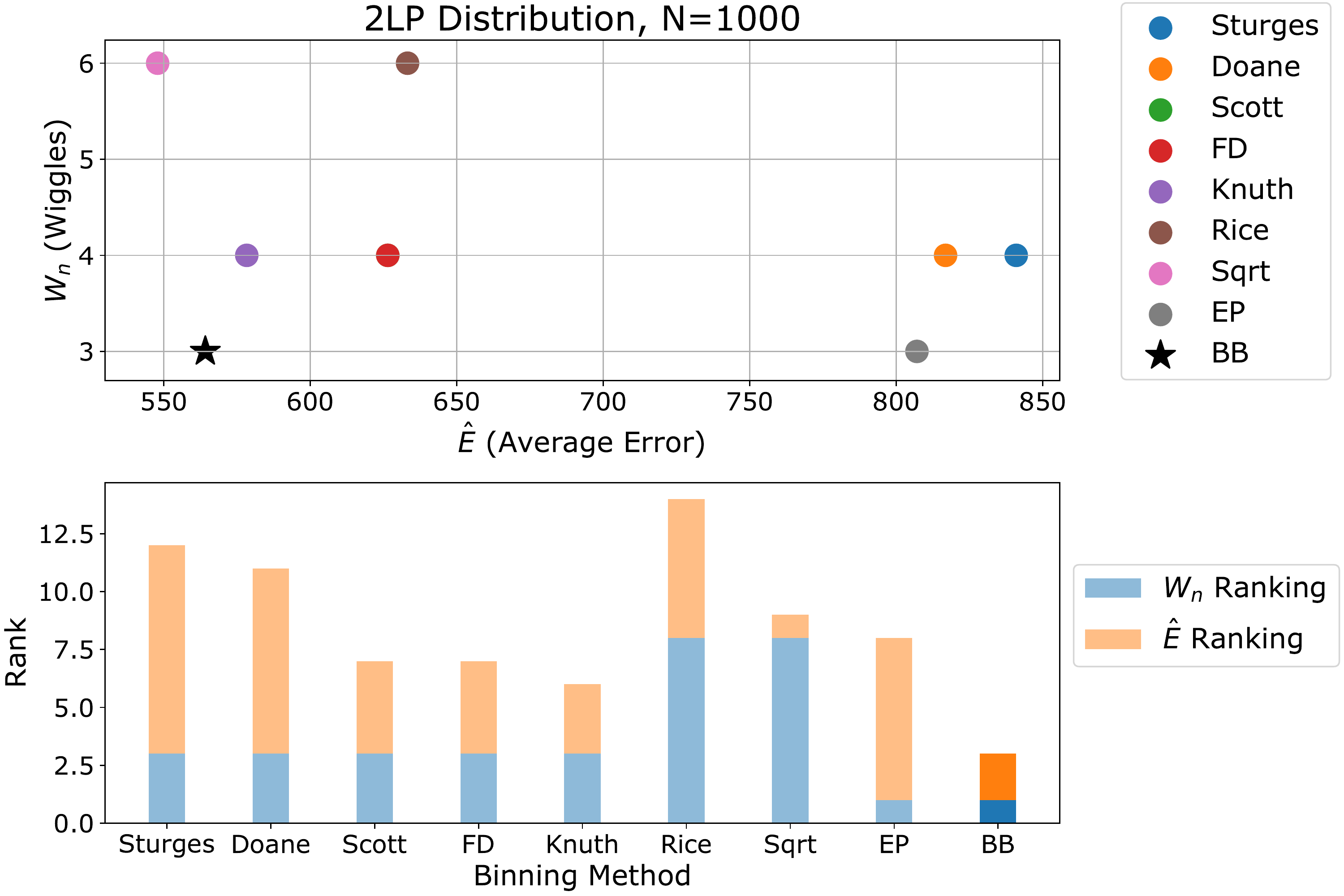}
        \vspace*{2mm}
    \end{subfigure}
    \begin{subfigure}{\columnwidth}
        \centering
        \includegraphics[width=0.9\columnwidth]{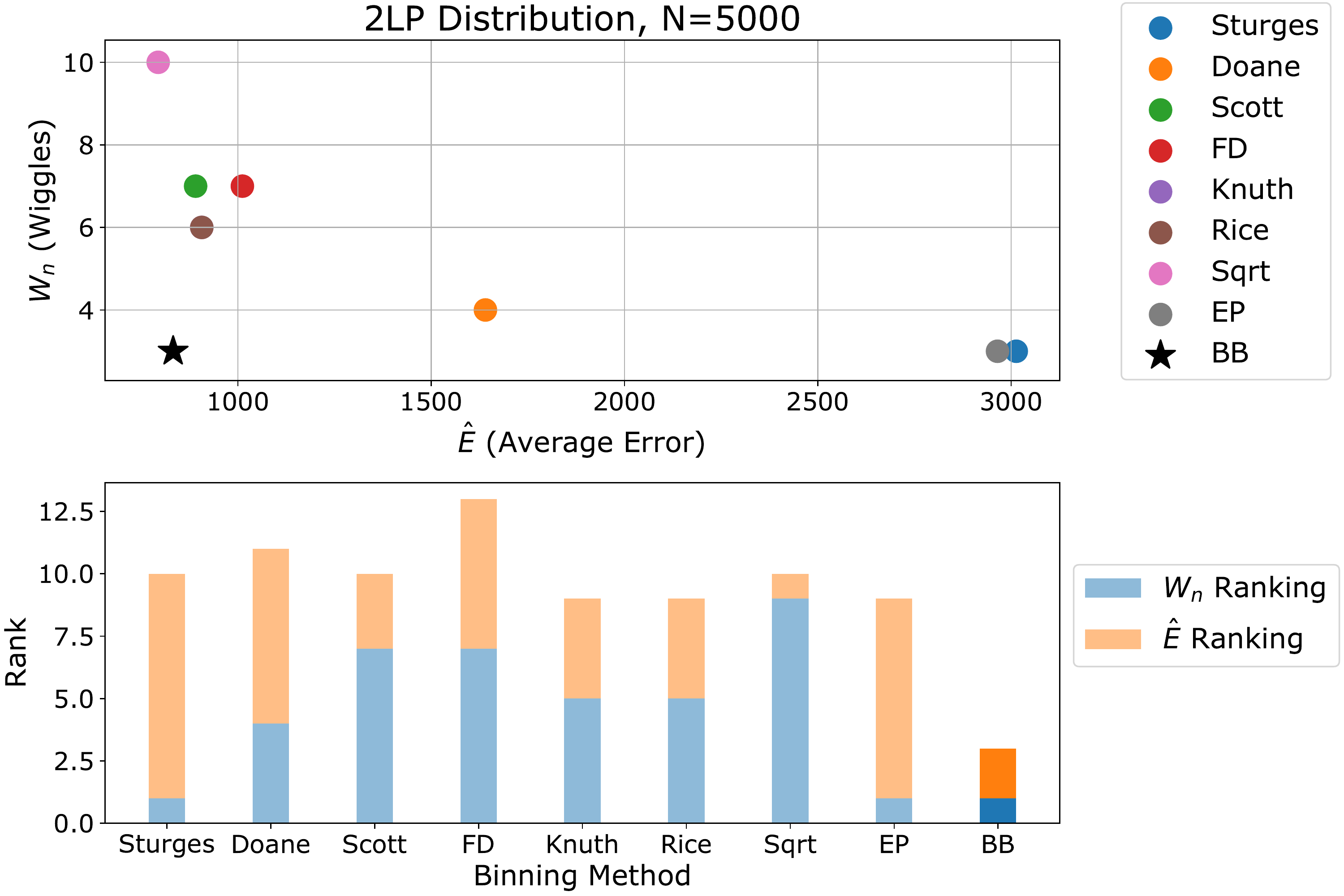}
        \vspace*{2mm}
    \end{subfigure}
    \begin{subfigure}{\columnwidth}
        \centering
        \includegraphics[width=0.9\columnwidth]{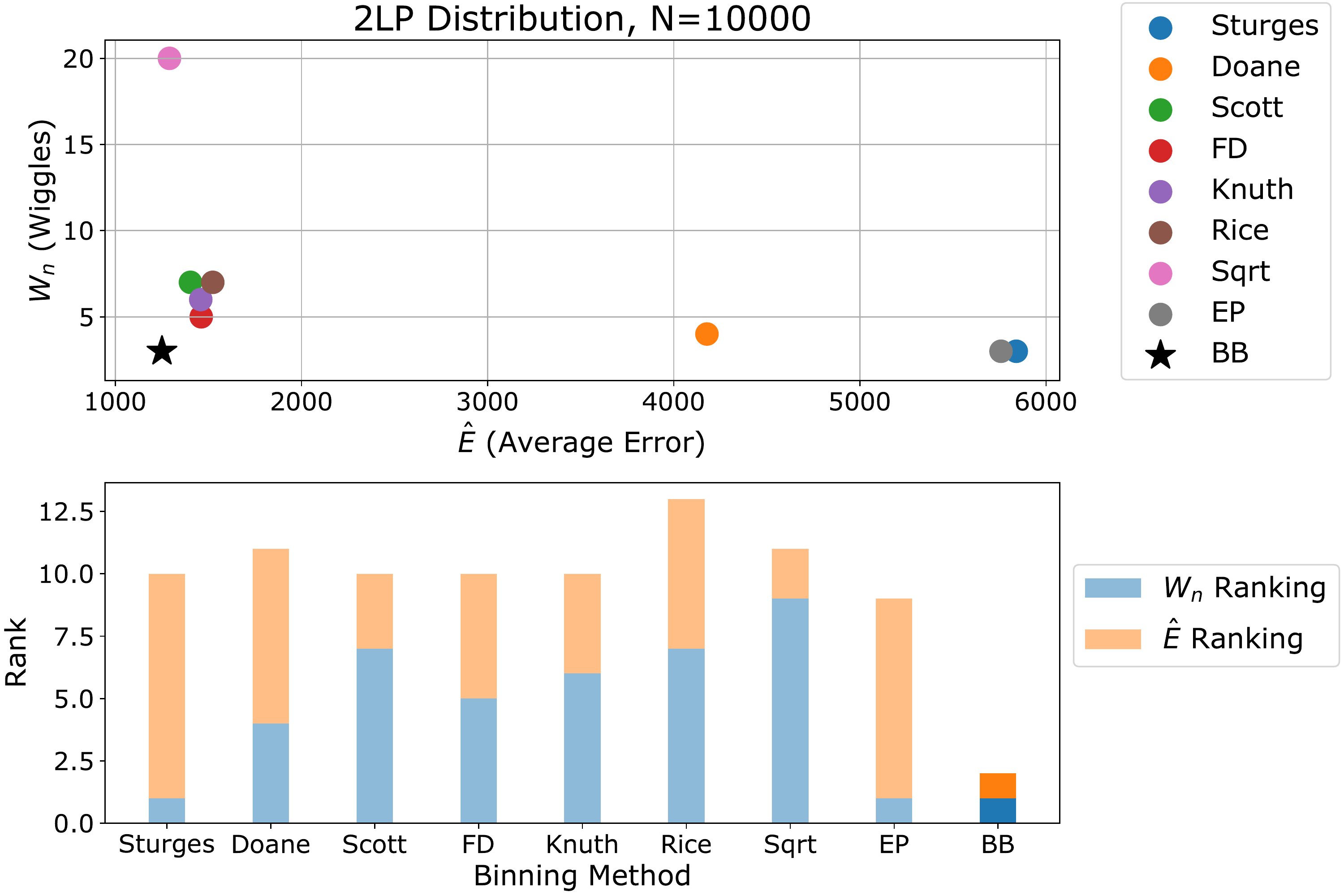}
        \vspace*{2mm}
    \end{subfigure}
    \caption{Metric values and combined ranks for a bimodal distribution (2LP) for different sized datasets.\label{fig:metric_2LP}}
\end{figure*}
\begin{figure*}[!htb]
    \centering
    \begin{subfigure}{\columnwidth}
        \centering
        \includegraphics[width=0.9\columnwidth]{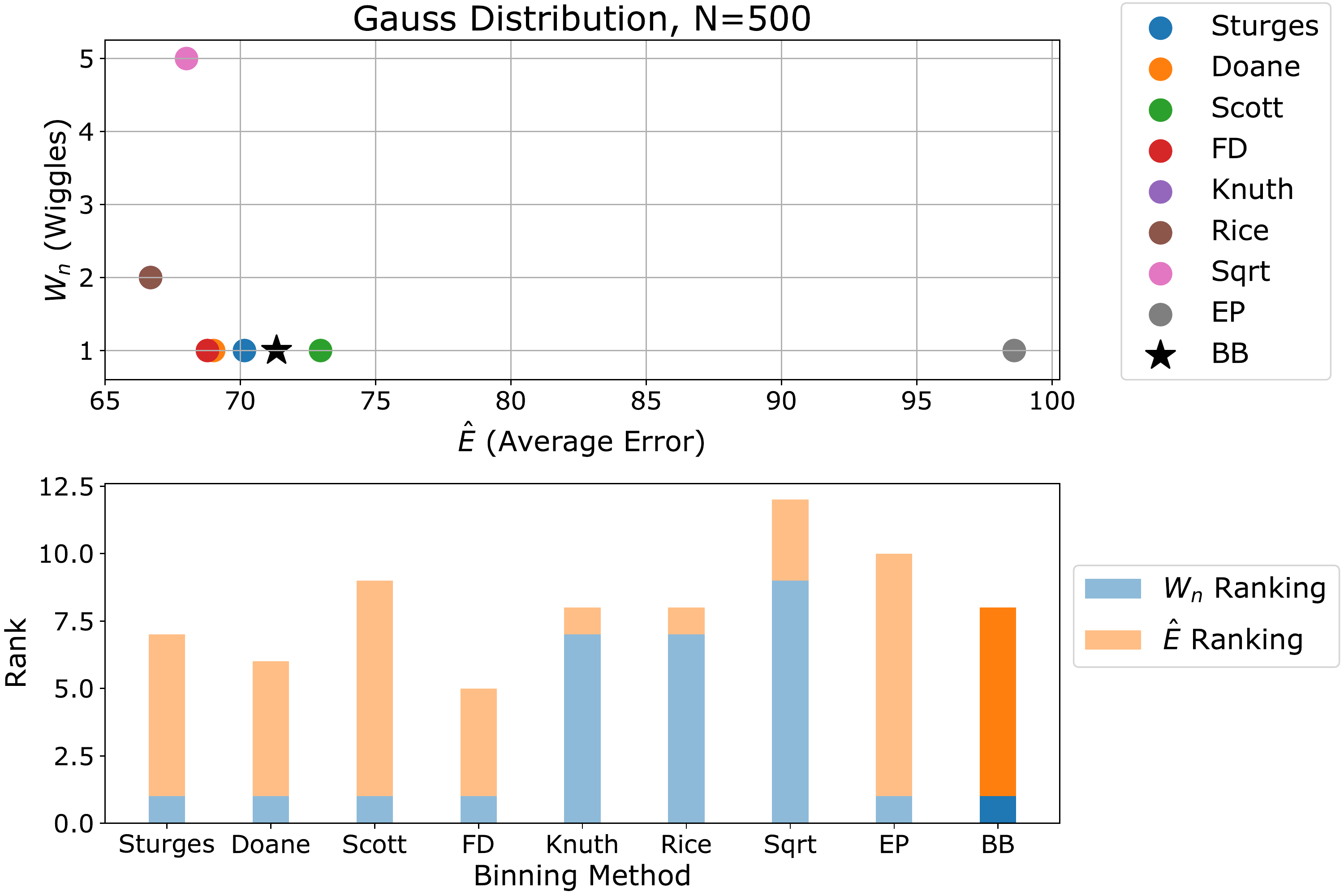}
        \vspace*{2mm}
    \end{subfigure}
    \begin{subfigure}{\columnwidth}
        \centering
        \includegraphics[width=0.9\columnwidth]{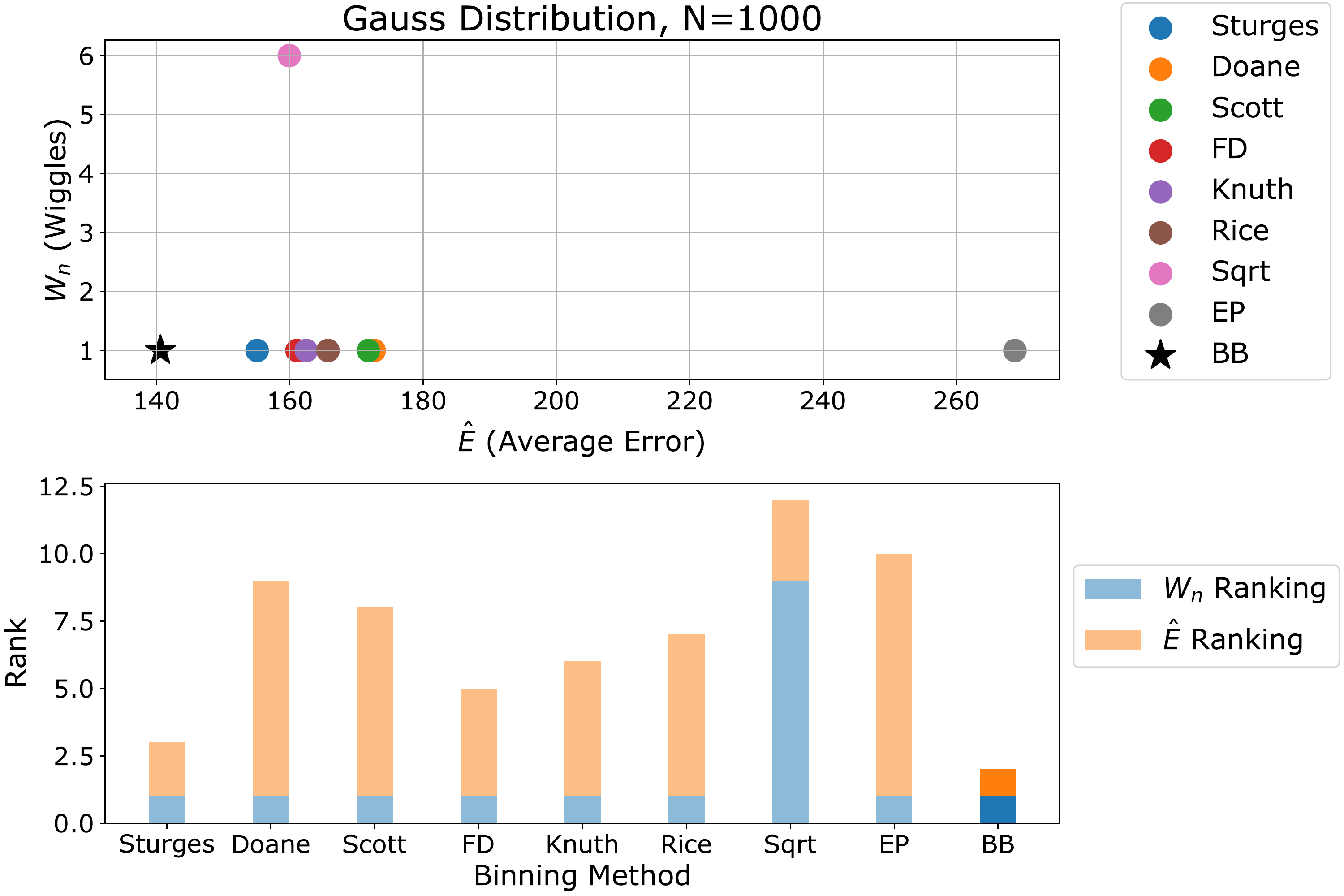}
        \vspace*{2mm}
    \end{subfigure}
    \begin{subfigure}{\columnwidth}
        \centering
        \includegraphics[width=0.9\columnwidth]{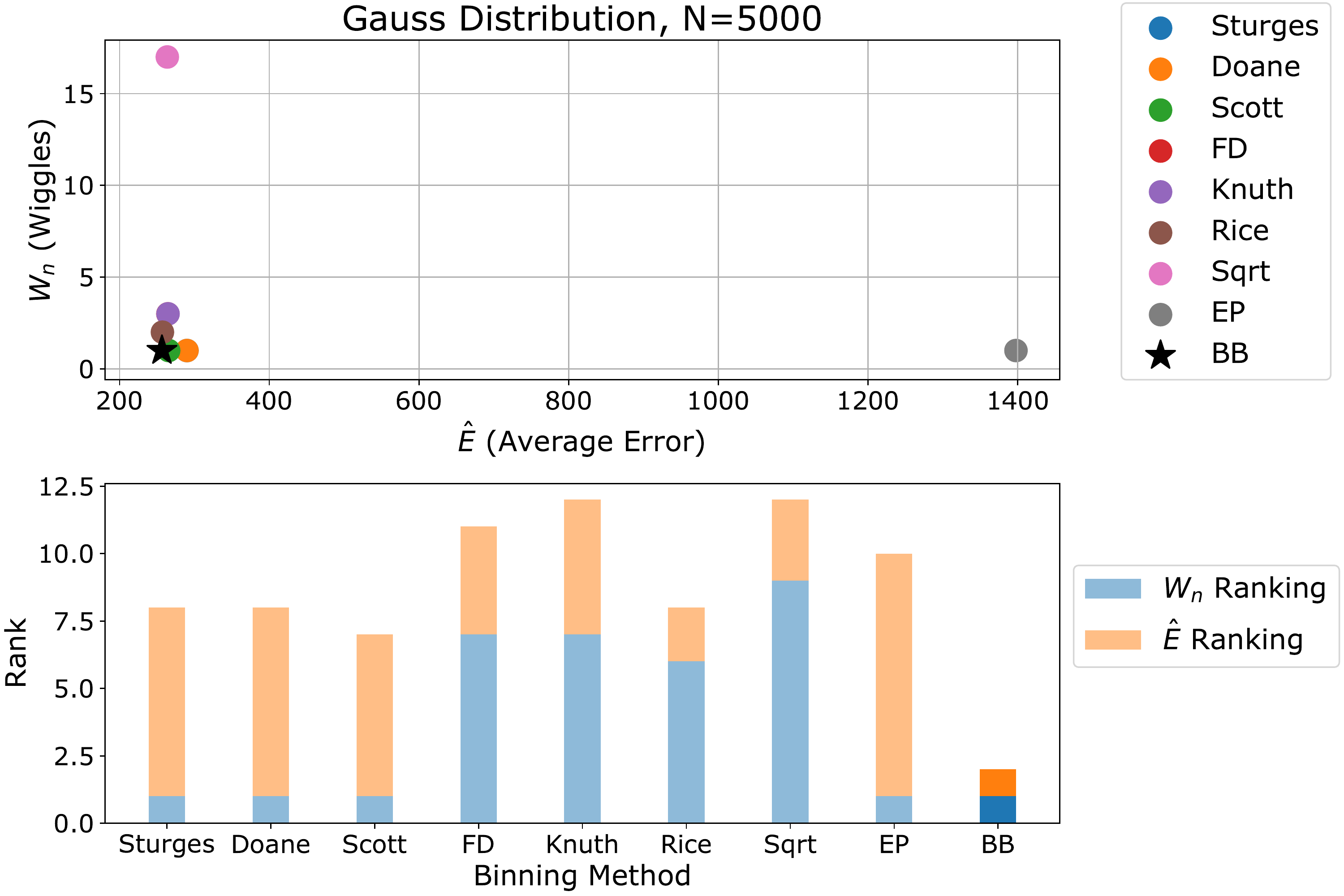}
        \vspace*{2mm}
    \end{subfigure}
    \begin{subfigure}{\columnwidth}
        \centering
        \includegraphics[width=0.9\columnwidth]{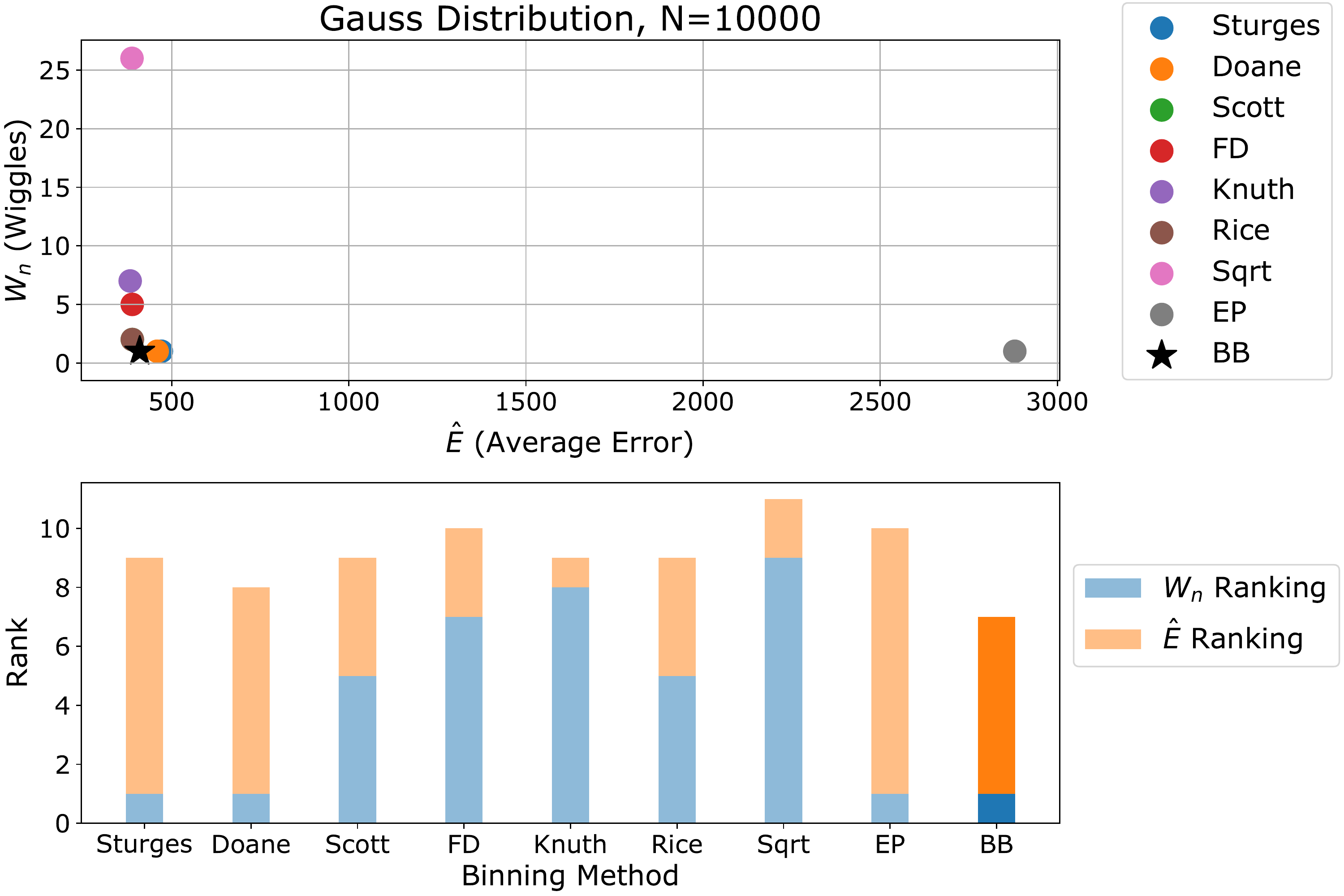}
        \vspace*{2mm}
    \end{subfigure}
    \caption{Metric values and combined ranks for Gaussian distribution (Gauss) for different sized datasets.\label{fig:metric_Gauss}}
\end{figure*}

The histograms for the DY, MuPT, jPT, 2LP, and Gauss distributions are shown in
Figs.~\ref{fig:hist_DY}-\ref{fig:hist_Gauss}. Plots of the two metrics, along with bar
charts of the combined ranks of the metrics are shown in
Figs.~\ref{fig:metric_DY}-\ref{fig:metric_Gauss}.
\par
For the majority of the examples shown here, Bayesian Blocks outperforms the other binning methods
in the context of the combined metric ranks.  In almost every scenario, Bayesian Blocks is minimally
wiggly, as the bin-to-bin statistical fluctuations are mitigated by the choice of bin width and
edges.  Infrequently, the Bayesian Block binning can produce a single ``spike'', arising from a very
narrow bin, which typically only occurs for relatively small datasets (see Fig.\ref{fig:hist_MuPT}
for $N$=500). This can be removed by tuning the prior, but at the cost of potentially increasing the
overall coarseness of the binning.  Typically, as the number of data points increases, the average
error metric of Bayesian Blocks becomes competitive or surpasses the values associated with
finer-binning methods.  For some cases, notably the jPT and 2LP scenarios, Bayesian Blocks quickly
becomes the objectively superior binning scheme with respect to both metrics. In general, as the
size of the dataset increases, Bayesian Blocks approaches or arrives at the minima of $\hat{E}$
(with respect to other methods), without any signifiant increase in $W_n$. The decision of which
binning heuristic to use for cases in which there is a relatively small dataset would depend on the
personal preference of the user.  

\section{\label{sec:conclusion}Summary and Conclusions}
\par
We have shown that histograms specified with the Bayesian Blocks algorithm have certain advantages
over typical histograms in high energy physics.  First, the flexible binning allows for a better
balance of statistical precision across a spectrum: sparse parts of distributions automatically have
larger bins, and dense parts have smaller bins.  Plots comparing two very similar distributions,
especially the ratio of two similar distributions, are cleaner and easier to interpret, as
illustrated by the plots in Figs.~\ref{fig:dimuon} and~\ref{fig:jet}.
\par
The benefits of the Bayesian Block algorithm are investigated quantitatively with respect to
other popular binning schemes. We defined two metrics, $W_n$ and $\hat{E}$, to describe the visual
appeal and the modeling accuracy of a given histogram.  Given these metrics, and a host of
distributions inspired by common HEP scenarios, the Bayesian Blocks algorithm proves to be a
competitive or superior candidate, especially as size of the dataset increases.  The binning it
produces is extremely robust to statistical fluctuations, without sacrificing modeling accuracy.
This makes Bayesian Blocks and ideal candidate for displaying and exploring data, without misleading
the viewer with spurious bin-to-bin fluctuations or obscuring sharp features with excessively coarse
binning.
\par
This paper serves to introduce the Bayesian Blocks algorithm to the
high energy physics community.  Theoretical background can be found
in the references, and applications await further data analysis elsewhere. The Bayesian Block
implementation used in this paper can be found in the Scikit-HEP~\cite{Scikit-HEP} python package.

\begin{acknowledgements}
    We thank the CPC journal referees for their feedback on the original version of this paper, and
    to Ryan Pellico for his assistance with mathematical inquires.  The work reported in this
    article was funded by grant DE-SC0015910 from the U.S. Department of Energy.
\end{acknowledgements}

\Urlmuskip=0mu plus 1mu\relax
\bibliography{main}

\end{document}